# Deepfakes Generation and Detection: State-of-the-art, open challenges, countermeasures, and way forward

**Momina Masood[1], Mariam Nawaz[2], Khalid Mahmood Malik[3], Ali Javed[4], Aun Irtaza[5]**
[1,2]Department of Computer Science, University of Engineering and Technology-Taxila, Pakistan
[3,4]Department of Computer Science and Engineering, Oakland University, Rochester, MI, USA
[5]Electrical and Computer Engineering Department, University of Michigan-Dearborn, MI, USA

**Abstract**
Easy access to audio-visual content on social media, combined with the availability of modern tools such as Tensorflow or Keras, open-source trained models, and economical computing infrastructure, and the rapid evolution of deep-learning (DL) methods, especially Generative Adversarial Networks (GAN), have made it possible to generate deepfakes to disseminate disinformation, revenge porn, financial frauds, hoaxes, and to disrupt government functioning. The existing surveys have mainly focused on the detection of deepfake images and videos. This paper provides a comprehensive review and detailed analysis of existing tools and machine learning (ML) based approaches for deepfake generation and the methodologies used to detect such manipulations for both audio and visual deepfakes. For each category of deepfake, we discuss information related to manipulation approaches, current public datasets, and key standards for the performance evaluation of deepfake detection techniques along with their results. Additionally, we also discuss open challenges and enumerate future directions to guide future researchers on issues that need to be considered to improve the domains of both deepfake generation and detection. This work is expected to assist the readers in understanding the creation and detection mechanisms of deepfakes, along with their current limitations and future direction.

## 1 Introduction

The availability of economical digital smart devices like cellphones, tablets, laptops, and digital cameras has resulted in the exponential growth of multimedia content (e.g. images and videos) in cyberspace. Additionally, the evolution of social media over the last decade has allowed people to share captured multimedia content rapidly, leading to a significant increase in multimedia content generation and ease of access to it. At the same time, we have witnessed tremendous advancement in the field of ML with the introduction of sophisticated algorithms that can easily manipulate multimedia content to spread disinformation online through social media platforms. Given the ease with which false information may be created and spread, it has become increasingly difficult to know the truth and trust the information, which may result in harmful consequences. Moreover, today we live in a "post-truth" era, where a piece of information or disinformation is utilized by malevolent actors to manipulate public opinion. Disinformation is an active measure that has the potential to cause severe damage: election manipulation, creation of warmongering situations, defaming any person, etc. In recent times, the deepfakes generation has significantly advanced and it could be used to propagate disinformation around the globe and may pose a severe threat, in the form of fake news, in the future. Deepfakes are synthesized, AI-generated, videos, and audio. The use of videos as evidence in every sector of litigation and criminal justice proceedings is currently the norm. A video admitted as evidence must be authentic and its integrity must be verified. On the other hand, most of the existing multimedia forensic examiners face the challenge of analyzing as evidence multimedia files that originate from social networks and sharing websites, e.g., YouTube, Facebook, etc. Satisfying the authentication and integrity requirements and flagging the manipulated videos on social media is a challenging task, especially as deepfakes generation becomes more sophisticated. Once the deepfakes have been created, the further use of powerful, sophisticated, and easy-to-use manipulation tools (e.g. Zao[1], REFACE[2], FaceApp[3], Audacity [4], Soundforge [5]) could make authentication and integrity verification of generated videos even more difficult task.

Deepfakes videos can be categorized into the following types: i) face-swap ii) lip-synching iii) puppet-master iv) face synthesis and attribute manipulation, and v) audio deepfakes. In face-swap deepfakes, the face of the source person is replaced with the target person to generate a fake video of the target person, trying to portray actions to the target person which in reality the source person has done. Face-swap-oriented deepfakes are usually generated to target the popularity or reputation of famous personalities by showing them in scenarios in which they never appeared [6], and to damage reputations in the face of the public, for example, in non-consensual pornography [7]. In lip-synching-based deepfakes, the movements of the target person's lips are transformed to make them consistent with some specific audio recording. Lip-syncing is generated with the aim of showing an individual speaking in a way in which the attacker devises the victim to speak. With puppet-master, deepfakes are created by mimicking the expressions of the target person, such as eye movement, facial expressions, and head movement. Puppet-master deepfakes aim to hijack the source person's expression, or even the full-body, [8] in a video, and to animate according to the impersonator's desire. Face synthesis and attribute manipulation involve the generation of photo-realistic face images and facial attribute editing. This manipulation is generated to spread disinformation on social media using fake profiles. Lastly, audio deepfakes focus on the generation of the target speaker's voice using deep learning techniques to portray the speaker saying something they have not said [9, 10]. The fake voices can be generated using either text-to-speech synthesis (TTS) or voice conversion (VC).

Unlike deepfake videos, less attention has been paid to the detection of audio deepfakes. In the last few years, voice manipulation has also become very sophisticated. Synthetic voices are not only a threat to automated speaker verification systems, but also to voice-controlled systems deployed in the Internet of Things (IoT) settings [11, 12]. Voice cloning has tremendous potential to destroy public trust and to empower criminals to manipulate business dealings or private phone calls. For example, recently a case was reported in which bank robbers used voice cloning of a company executive's speech to dupe their subordinates into transferring hundreds of thousands of dollars into a secret account [13]. The integration of voice cloning into deepfakes is expected to become a unique challenge for deepfake detection. Therefore, it is important that, unlike current approaches that focus only on detecting video signal manipulations, audio forgeries should also be examined.

There are no existing recently published surveys on deepfake generation and detection that focus on the generation and detection of both the audio and video modalities of deepfakes. Most of the existing surveys focus only on reviewing deepfakes images, and video detection. In [14], the main focus was on generic image manipulation and multimedia forensic techniques. However, this work has not discussed deepfake generation techniques. In [15], an overview of face manipulation and detection techniques was presented. Another survey [16] was presented on visual deepfakes detection approaches but does not discuss audio cloning and its detection. The latest work presented by Mirsky et al. [17] gives an in-depth analysis of visual deepfake creation techniques, however, deepfake detection approaches are only briefly discussed. Moreover, this work [17] lacks a discussion of audio deepfakes. According to the best of our knowledge, this paper is the first attempt to provide a detailed analysis and review of both the audio and visual deepfake detection techniques, as well as generative approaches. The following are the main contributions of our work:

i. To give the research community an insight into various types of video and audio-based deepfake generation and detection methods.
ii. To provide the reader with the latest improvements, trends, limitations, and challenges in the field of audio-visual deepfakes.
iii. To give an understanding to the reader about the possible implications of audio-visual deepfakes.
iv. To act as a guide to the reader to understand the future trends of audio and visual deepfakes.

The rest of the paper is organized as follows. Section 2 presents a discussion of deepfakes as a source of disinformation. In Section 3, the history and evolution of deepfakes is briefly discussed. Section 4 presents the overview of state-of-the-art audio and visual deepfake generation and detection techniques. We have also discussed open challenges for both audio-visual deepfake generation and detection in Section 4. Section 5 presents the available datasets used for both audio and video deepfakes detection. In Section 6, we discuss the possible future trends of both deepfakes generation and detection, and finally, we conclude our work in Section 7.

**Methodology**

In this paper, we have reviewed the existing approaches, used for the generation and detection of audio and visual manipulations, published in various authentic venues. A detailed description of the approach and protocol employed for the review is given in Table 1.

Table 1: Literature collection and preparation protocol

| Preparation Protocol | Description |
|---|---|
| Purpose | • To provide a brief overview of existing state-of-the-art techniques and identify potential gaps both in audio-visual deepfake generation and detection.<br>• To provide systematic review and structure to the existing state-of-the-art techniques with respect to each category of audio-visual deep fake generation and detection. |
| Data sources | Google Scholar, Springer Link, ACM digital library, IEEE explorer and DBLP |
| Query | Methodological approach was designed on data sources mentioned above and the following query strings were used:<br>Deepfakes/Faceswap/ Face reenactment/ lip-syncing /Deepfakes AND Faceswap/ Deepfakes AND Face reenactment/ Deepfakes AND lip-syncing/ GAN synthesized/ face manipulation/ Attribute Manipulation/GAN AND Puppet Mastery/ GAN AND Expression Manipulation/ Video Synthesis/ Audio synthesis/ Deep learning AND TTS/ Deep learning AND Voice Conversion/ Deep learning AND Voice Cloning/Deepfakes AND Dataset/ Deepfakes AND Audio/ Deepfakes AND Video/Deepfakes AND image |
| Method | We have systematically categorized the literature of video and audio deepfakes as follows (see figure 1):<br>a) Video deepfake generation and detection into the following categories: face swap, lip-syncing, puppet-mastery, entire face synthesis, and facial attribute manipulation.<br>b) Audio deepfake generation and detection into the following categories: text-to-speech synthesis and voice conversion. |
| Size | A total of 350 papers were retrieved using the method and query mentioned above from listed data sources till 07-08-2021. We have selected only those studies that were relevant and passed the criteria to be 'deepfakes' were included in the positive set. The other relevant but not in the positive set were included in the negative set. All other studies were excluded from the final selected papers i.e., white paper and articles. |
| Study types/inclusion and exclusion | The peer-reviewed journal papers, and articles of conference proceedings, were given more importance. Additionally, few articles from archive literature were also considered. |

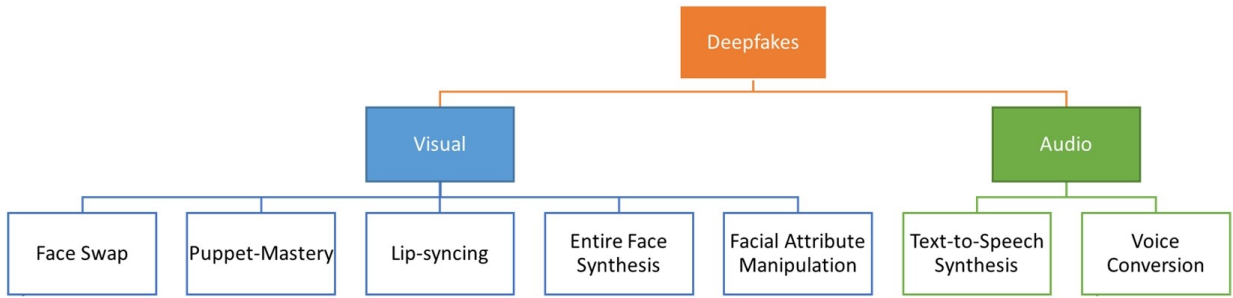

Figure 1: Categorization of Audio and Visual Deepfakes

## 2 Disinformation and Misinformation using Deepfakes

Misinformation is defined as false or inaccurate information that is communicated, regardless of an intention to deceive, whereas disinformation is the set of strategies employed by influential society representatives to fabricate original information to achieve the planned political or financial objectives. It is expected to become the main process of intentionally spreading manipulated news to affect public opinion or obscure reality. Because of the extensive use of social media platforms, it is now very easy to spread false news [18]. Although all categories of fake multimedia (i.e. fake news, fake images, and fake audio) could be sources of disinformation and misinformation, audiovisual-based deepfakes are expected to be much more devastating. Historically, deepfakes were created to make famous personalities controversial among their fans. For example, in 2017 a celebrity faced such a situation when a fake pornographic video was circulated in cyberspace. This is evidence that deepfakes can be used to damage the reputations, i.e. character's assassination of renowned people to defame them [16], blackmailing individuals for monetary benefits, or to create political or religious unrest by targeting politicians or religious scholars with fake videos/speeches [19], etc. This damage is not limited to targeting individuals; rather deepfakes can be used to manipulate elections, create warmongering situations by showing fake videos of missiles launched to destroy the enemy state or used to deceive military analysts by portraying fake information, like showing a fake bridge across the river, to mislead troop deployment, and so on.

The deepfakes are expected to advance the following current sources of disinformation and misinformation to the next level.

**Trolls:** Independent Trolls are hobbyists who spread inflammatory information to cause disorder and reactions in society by playing with the emotions of people [20]. For example, posting audio-visual manipulated racist or sexist content and infuriating individuals may promote hatred among the individuals. Similarly, during the 2020 election campaign of US President Donald Trump, conflicting narratives about Trump and Biden were circulated on social media, contributing to an environment of fear [21]. As opposed to independent trolls who spread false information for their own satisfaction, hired trolls will perform the same job for monetary benefits. Different actors, like political parties, businessmen, and companies routinely hire people to forge news related to their competitors and spread it in the market [22]. For example, according to a report published by Western intelligence [23], Russia is running "troll farms," where trolls are trained to affect conversations related to national or international issues. According to these reports, deepfake videos generated by hired trolls are the newest weapon in the ongoing fabricated news war that can bring a more devastating effect on society.

**Bots:** Bots are automated software or algorithms used to spread fabricated or misleading content among people. A study published in [24, 25] concludes that during the US election campaign-2016, bots were employed to generate one-fifth of the tweets during the last month of the campaign. The emergence of deepfakes has empowered the negative impact of bots i.e. recently, a messaging app named telegram [26] used bots to produce nude pictures of women, which is under investigation by Italian authorities.

**Conspiracy Theorists:** Conspiracy Theorists can range from nonprofessional filmmakers to Reddit agents who spread vague and doubtful claims on the internet either through "documentaries" or by posting stories and memes [27]. They believe that certain prominent communities are running the public while concealing their activities, like conspiracy theories about a Jewish plan to control the world [27, 28]. Moreover, recently, several conspiracy theorists have connected the current COVID pandemic with the USA and China. In such a situation, the use of fabricated audio-visual deepfake content by these theorists can increase controversy in global politics.

**Hyper-partisan Media:** Hyper-partisan media includes fake news websites and blogs which intentionally spread false information. Because of the extensive usage of social media, Hyper-partisan media is one of the biggest potential incubators for spreading fabricated news among the people [29]. The convincing AI-generated fake content can assist these bloggers to easily spread disinformation to attract visitors or increase views. As social platforms are largely independent and ad-driven mediums, spreading fabricated information may purely be a profit-making strategy.

**Politicians:** One of the main sources of disinformation is the political parties themselves, which may spread manipulated information for point-scoring. Due to a large number of followers on social platforms, politicians are central nodes in online networks. So, politicians can use their fame and public support to spread false news among their followers. To defame opponent parties, politicians can use deepfakes to post controversial content about their competitors on conventional media [27].

**Foreign Governments:** As the Internet has converted the world into a "Global Village," it is easy for conflicting countries to spread false news to advance their agendas abroad. Their motive is to target the reputation of a country in the rest of the world. Many countries are running government-sponsored social media accounts, websites, and applications, contributing to political propaganda globally. Particularly, the governments of China, Israel, Turkey, Russia, the UK, Ukraine, India, and North Korea, etc. are believed to be involved in using 'digital footsoldiers' to smear opponents, spreading disinformation, and posting fake texts for 'pocket money' [30]. These countries run numerous official social sites over various online platforms like Twitter, Instagram, and Facebook, etc. [31]. The ability to doctor multimedia content has become so easy that private actors may be able to initiate foreign attacks on their own to increase the stress among countries.

## 3 DeepFakes Evolution

The earliest example of manipulated multimedia content occurred in 1860 when a portrait of southern politician John Calhoun was skillfully manipulated by replacing his head with that of US President Abraham Lincoln [32]. Usually, such manipulation is accomplished by adding (splicing), removing (inpainting), and replicating (copy-move) the objects within or between two images [14]. Then, suitable post-processing steps like scaling, rotating, and color adjustment are applied to improve the visual appearance, scale, and perspective coherence.

Aside from these traditional manipulation methods, advancements in Computer Graphics and DL techniques now offer a variety of different automated approaches for digital manipulation with better semantic consistency. The recent trend involves the synthesis of videos from scratch using autoencoders, or Generative Adversarial Network (GAN), for different applications [33] and, more specifically, photorealistic human face generation based on any attribute [34-37]. Another pervasive manipulation, called "shallow fakes" or "cheap fakes," are audio-visual manipulations created using cheaper and more accessible software. Shallow fakes involve basic editing of a video utilizing slowing, speeding, cutting, and selectively splicing together unaltered existing footage that can alter the whole context of the information delivered. In May 2019, a video of US Speaker Nancy Pelosi was selectively edited to make it appear that she was slurring her words and was drunk or confused [38]. The video was shared on Facebook and received more than 2.2 million views within 48 hours. Video manipulation for the entertainment industry, specifically in film production, has been done for decades. Fig. 2 shows the evolution of deepfakes over the years. An early notable academic project was Video Rewrite Program [39], intended for applications in movie dubbing, published in 1997. It was the first software used to automatically reanimate facial movements in an existing video to a different audio track, and it achieved surprisingly convincing results.

The first true deepfake appeared online in September 2017 when a Reddit user named "deepfake" posted a series of computer-generated videos of famous actresses with their faces swapped onto pornographic content [16]. Another notorious deepfake case was the release of the deepNude application that allowed users to generate fake nude images [40]. This was the beginning of when deepfakes gained wider recognition within a large community. Today deepfake technology/applications, i.e. FakeApp [41], FaceSwap [42], and ZAO [1] are easily accessible, and users without a computer engineering background can create a fake video within seconds. Moreover, open-source projects on GitHub, such as DeepFaceLab [43] and related tutorials, are easily available on YouTube. A list of other available deepfake creation applications, software, and open-source projects is given in Table 2. Contemporary academic projects that lead to the development of deepfake technology are Face2Face [36] and Synthesizing Obama [35], published in 2016 and 2017 respectively. Face2Face [36] captures the real-time facial expressions of the source person as they talk into a commodity webcam. It modifies the target person's face in the original video to depict them, mimicking source facial expressions. Synthesizing Obama [35] is a video rewrite 2.0 program, used to modify the mouth movement in the video footage of a person to depict the person saying the words contained in an arbitrary audio clip. These works [35, 36] are focused on the manipulation of the head and facial region only. Recent development expands the application of deepfakes to the entire body, [8, 44] and the generation of deepfakes from a single image [45-47].

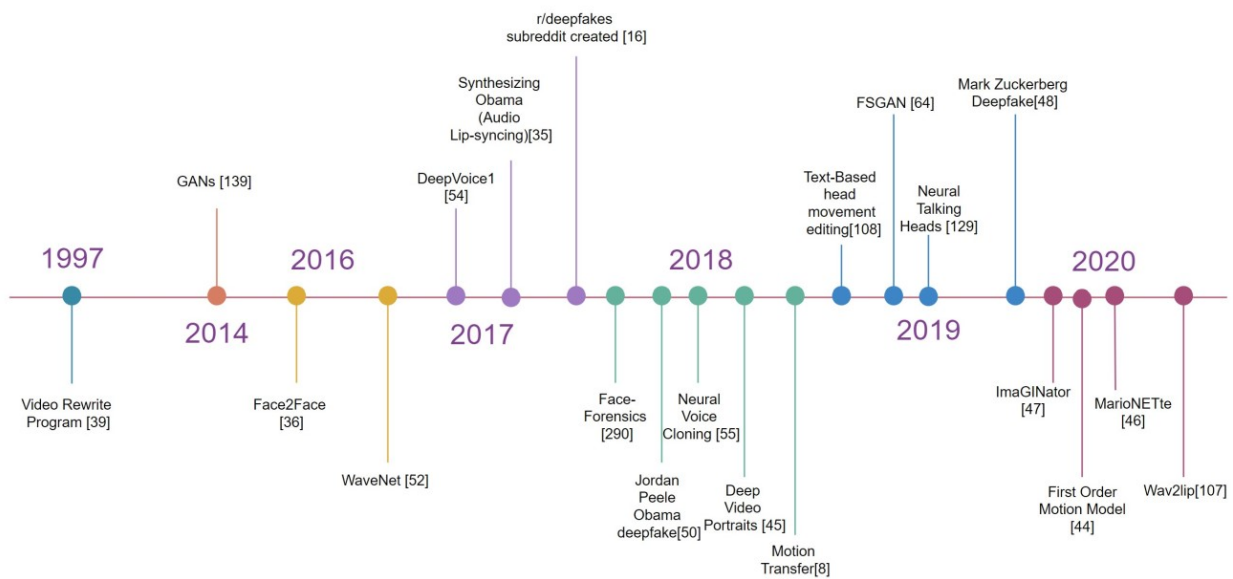

Figure 2: The timeline of Deepfakes evolution

Most deepfakes currently present on social platforms like YouTube, Facebook or Twitter may be regarded as harmless, entertaining, or artistic. However, there are also some examples where deepfakes have been used for revenge porn, hoaxes, political or non-political influence, and financial fraud [48, 49]. In 2018, a deepfake video went viral online in which former U.S. President Barak Obama appeared to insult the current president, Donald Trump [50]. In June 2019, a fake video of Facebook CEO Mark Zuckerberg was posted to Instagram by the Israeli advertising company "Canny" [48]. Recently, extremely realistic deepfake videos of Tom Cruise posted on the TikTok platform have gained 1.4million views within a few days [51].

**Table 2: An overview of Audio-visual deepfakes generation software, applications, and open-source projects**

| Tool | Type | Reference/Developer | Technique |
|---|---|---|---|
| **Cheap fakes** | | | |
| Adobe Premiere | Commercial Desktop Software | Adobe | Audio Video Editing, AI-powered video reframing |
| Corel VideoStudio | Commercial Desktop Software | Corel | Proprietary AI |
| **Lip-sync** | | | |
| dynalips | Commercial Web App | www.dynalips.com/ | Proprietary |
| crazytalk | Commercial Web App | www.reallusion.com/crazytalk/ | Proprietary |
| Wav2Lip | Open source implementation | github.com/Rudrabha/Wav2Lip | GAN with pre-trained discriminator network and visual quality loss function |
| **Facial Attribute Manipulation** | | | |
| FaceApp | MobileApp | FaceApp Inc | Deep generative CNNs |
| Adobe | Commercial Desktop Software | Adobe | DNNs + filters |
| Rosebud | Commercial Web App | www.rosebud.ai/ | Proprietary AI |
| **Face Swap** | | | |
| ZAO | Mobile app | Momo Inc | Proprietary |
| REFACE | Mobile app | Neocortext, Inc | Proprietary |
| Reflect | Mobile app | Neocortext, Inc | Proprietary |
| Impressions | Mobile app | Synthesized Media, Inc. | Proprietary |
| FakeApp | Desktop App | www.malavida.com/en/soft/fakeapp/ | GAN |
| FaceSwap | Open source implementation | faceswapweb.com/ | Employed two pairs of encoder-decoder. Shared encoder parameters. |
| DFaker | Open source implementation | github.com/dfaker/df | For face reconstruction DSSIM loss function [34] is utilized. Keras library-based implementation. |
| DeepFaceLab | Open source implementation | github.com/iperov/DeepFaceLab | - provide several face extraction methods, e.g. dlib, MTCNN, S3FD etc.<br>- Extend different Faceswap model i.e. H64, H128, LIAEF128, SAE [33]. |
| FaceSwapGAN | Open source implementation | github.com/shaoanlu/faceswap-GAN | Uses two loss functions namely adversarial loss and perceptual loss to the auto-encoder. |
| DeepFake-tf | Open source implementation | github.com/StromWine/DeepFake-tf | Same as DFaker however, used tensor-flow for implementation. |
| Faceswapweb | Commercial Web App | faceswapweb.com/ | GAN |
| **Face Reenactment** | | | |
| Face2Face | Open source implementation | web.stanford.edu/~zollhoef/papers/CVPR2016_Face2Face/page.html | Uses 3DMM and ML technique |
| Dynamixyz | Commercial Desktop Software | www.dynamixyz.com/ | Machine-learning |
| FaceIT3 | Open source implementation | github.com/alew3/faceit_live3 | GAN |
| **Face Generation** | | | |
| Generated Photos | Commercial Web App | generated.photos/ | StyleGAN |
| **Voice Synthesis** | | | |
| Overdub | Commercial Web App | www.descript.com/overdub | Proprietary (AI based) |
| Respeecher | Commercial Web App | www.respeecher.com/ | Combined traditional digital signal processing algorithms with proprietary deep generative modeling techniques |
| SV2TTS | Open source implementation | github.com/CorentinJ/Real-Time-Voice-Cloning | LSTM with Generalized end-to-end loss |
| ResembleAI | Commercial Web App | www.resemble.ai/ | Proprietary (AI based) |
| Voicery | Commercial Web App | www.voicery.com/ | Proprietary AI and deep learning |
| VoiceApp | Mobile app | Zoezi AB | Proprietary (AI-based) |

Apart from visual manipulation, audio deepfakes are a new form of cyber-attack, with the potential to cause severe damage to individuals due to highly sophisticated speech synthesis techniques i.e. WaveNet [52], Tacotron [53], and deep voice1 [54]. Fake audio-assisted financial scams have increased significantly in 2019 due to the progression in speech synthesis technology. In August 2019, a European company's chief executive officer, tricked by an audio deepfake, made a fraudulent transfer of $243,000 [13]. A voice-mimicking AI software was used to clone the voice patterns of the victim by training ML algorithms using audio recordings obtained from the internet. If such techniques can be used to imitate the voice of a top government official or a military leader and applied at scale, it could have serious national security implications [55].

# 4 Audio Visual Deepfakes Types and Categorization of Literature

This section provides an in-depth analysis of existing state-of-the-art methods for audio and visual deepfakes. A review for each category of deepfake in terms of creation and detection is provided to give a deeper understanding of the various approaches. We provide a critical investigation of existing literature which includes the technologies, their capabilities, limitations, challenges, and future trends for both deepfake creation and detection. Deepfakes are broadly categorized into two groups namely visual and audio manipulations depending on the targeted forged modality (Fig. 1). The visual deepfakes are further grouped into the following types based on manipulation level (i) face swap/identity swap, (ii) lip-syncing, (iii) face-reenactment/puppet-mastery, iv) entire face synthesis and v) facial attribute manipulation. The audio deepfakes are further classified as i) text-to-speech synthesis and ii) voice conversion.

## 4.1 Face-swap

**Generation**

Visual manipulation is nothing new; images and videos have been forged since the beginning. In face-swap [56], or face replacement, the face of the person in the source video is automatically replaced by the face in the target video, as shown in Fig. 3. Traditional face-swap approaches [57-59] generally take three steps to perform a face-swap operation. First, these tools detect the face in source images and then select a candidate face image from the facial library that is similar to input facial appearance and poses. Second, the method replaces the eyes, nose, and mouth of the face and further adjusts the lighting and color of the candidate face image to match the appearance of input images, and seamlessly blends the two faces. Finally, the third step ranks the blended candidate replacement by computing a match distance over the overlap region. These approaches may offer good results under certain conditions but have two major limitations. First, they completely replace the input face with the target face, and expressions of the input face image are lost. Second, the synthetic result is very rigid, and the replaced face looks unnatural e.g. it requires a matching pose to generate good results.

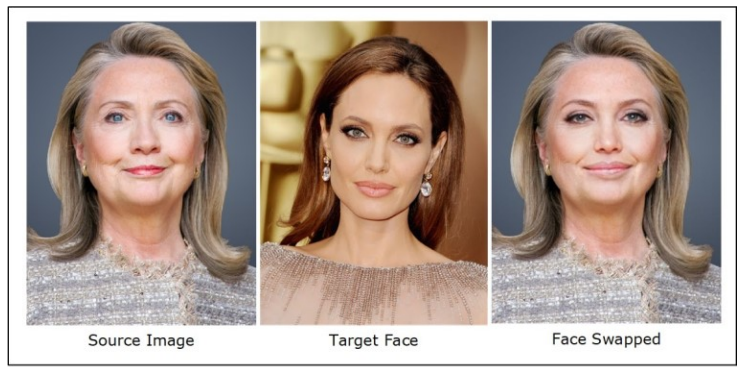

**Figure 3: A visual representation of Face-Swap based deepfake**

Recently, DL-based approaches have become popular for synthetic media creation due to their realistic results. At the same time, deepfakes showed how these approaches can be applied with automated digital multimedia manipulation. In 2017, the first deepfake video that appeared online was created using a face-swap approach, where the face of a celebrity was shown in pornographic content [16]. This approach used a neural network to morph a victim's face onto someone else's features while preserving the original facial expression. As time went on, face-swap software i.e. FakeApp [41] and FaceSwap [42] has made it both easier and quicker to produce deepfakes with more convincing results by replacing the face in a video. These approaches typically use two encoder-decoder pairs. Usually, an encoder is used to extract the latent features of a face from the image and then the decoder is used to reconstruct the face. To swap faces between the source and target image, two pairs of encoder and decoder are required, where each encoder is first trained on the source and then the target image. Once training is complete, the decoders are swapped, so that an original encoder of the source image and decoder of the target image is used to regenerate the target image with the features of the source image. The resulting image has the source's face on the target's face, while keeping the target's facial expressions. Fig. 4 is an example of a deepfake creation where the feature set of face A is connected with the decoder B to reconstruct face B from the original face A. The recently launched ZAO [1], REFACE [2], and FakeApp [41] applications are more popular due to their effectiveness in producing realistic face swap-based deepfakes. FakeApp allows the selective modification of facial parts. ZAO and REFACE have gone viral lately as less tech-savvy users can swap their faces with movie stars and embed themselves into well-known movies and TV clips. There are many publicly available implementations of face-swap technology using deep neural networks, such as FaceSwap [42], DFaker [60], DeepFaceLab [43], DeepFake-tf [61], and FaceSwapGAN [62], leading to the creation of a growing number of synthesized media clips.

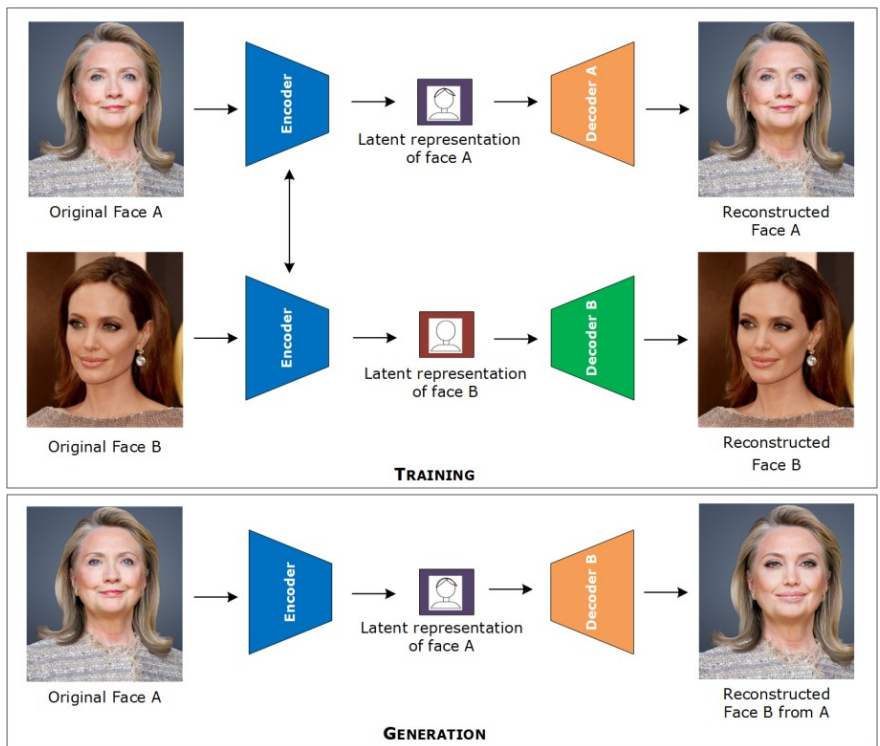

**Figure 4: Creation of a Deepfake using an auto-encoder and decoder. The same encoder-decoder pair is used to learn the latent features of the faces during training, while during generation decoders are swapped, such that latent face A is subjected to decoder B to generate face A with the features of face B**

Until recently, most of the research focused on advances in face-swapping technology, either using a reconstructed 3D morphable model (3DMM) [56, 63], or GANs based model [62, 64]. Korshunova et al. [63] proposed a convolution neural network (CNN) based approach that transferred the semantic content, e.g., face posture, facial expression, and illumination conditions of the input image to create that style in another image. They introduced a loss function that was a weighted combination of style loss, content loss, light loss, and total variation regularization. This method [63] generates more realistic deepfakes compared to [57], however, it requires a large amount of training data. Moreover, the trained model can be used to transform only one image at a time. Nirkin et. al [56] presented a method that used a full convolution network (FCN) for face segmentation and replacement while a 3DMM was established to estimate facial geometry and corresponding texture. Then the face reconstruction was performed on a target image by adjusting the model parameters. These approaches [56, 63] have the limitation of subject-specific or pair-specific training. Recently subject agnostic approaches have been proposed to address this limitation.

In [62], an improved deepfake generation approach using GAN was proposed which adds adversarial loss and perceptual loss to VGGface implemented in the auto-encoder architecture [42]. The addition of VGGFace perceptual loss made the direction of the eyeball appear more realistic and consistent with the input and also helped to smooth the artifacts added in the segmentation mask, resulting in a high-quality output video. FSGAN [64] allowed face swapping and reenactment in real-time by following the reenact and blend strategy. This method simultaneously manipulates pose, expression, and identity while producing high-quality and temporally coherent results. These GAN-based approaches [62, 64] outperform several existing autoencoder-decoder methods [41, 42] as they work without being explicitly trained on subject-specific images. Moreover, the iterative nature makes them well-suited for face manipulation tasks such as generating realistic images of fake faces.

Some of the work used a disentanglement concept for face swap by using VAEs. RSGAN [65] employed two separate VAEs to encode the latent representation of facial and hair regions respectively. Both encoders were conditioned to predict the attributes that describe the target identity. Another approach, FSNet [66], presented a framework to achieve face-swapping using a latent space, to separately encode the face region of the source identity and landmarks of the target identity, which was later combined to generate the swapped face. However, these approaches [65, 66] hardly preserves target attributes like target occlusion and illumination conditions.

Facial occlusions are always challenging to handle in face-swapping methods. In many cases, the facial region in the source or target is partially covered with hair, glasses, a hand, or some other object. This results in visual artifacts and inconsistencies in the resultant image. FaceShifter [67] generates a swapped face with high-fidelity and preserves the target attributes such as pose, expression, and occlusion. The last layer of a facial recognition classifier was used to encode the source identity and the target attributes, with feature maps being obtained via the U-Net decoder. These encoded features were passed to a novel generator with cascaded Adaptive Attentional Denormalization layers inside residual blocks which adaptively adjusted the identity region and target attributes. Finally, another network was used to fix occlusion inconsistencies and refine the results. Table 3 presents the detail of Face-swap based deepfakes creation approaches.

Table 3: An overview of Face-swap based Deepfake generation techniques

| Reference | Technique | Features | Dataset | Output Quality | Limitations |
|---|---|---|---|---|---|
| Faceswap [42] | Encoder-decoder | Facial landmarks | Private | 256×256 | ▪ Blurry results due to lossy compression<br>▪ Lack of pose, facial expression, gaze direction, hairstyle, and lighting<br>▪ Requires massive no. of target images |
| FaceSwapGAN [62] | GAN | VGGFace | VGGFace | 256×256 | ▪ Lack of texture details and generate overly smooth results |
| DeepFaceLab [68] | Encoder-decoder | Facial landmarks | Private | 256×256 | ▪ Fails to blend very different facial hues<br>▪ Requires target training data |
| Fast Face-swap [63] | CNN | VGGFace | ▪ CelebA (200,000 images)<br>▪ Yale Face Database B (different pose and lighting conditions) | 256×256 | ▪ Works for a single person only<br>▪ Gives better result for frontal face view<br>▪ Lack of skin texture details, e.g., smooth results and Facial Expression transfer<br>▪ Lack of occluding objects i.e. glasses |
| Nirkin et al. [56] | FCN-8s-VGG architecture | ▪ Basel Face Model to represent faces<br>▪ 3DDFA model for expression | IARPA Janus CS2 (1275 face videos) | 256×256 | ▪ Poor results in case of different image resolutions<br>▪ Fails to blend very different facial hues |
| Chen et al. [69] | VGG-16 net | 68 facial landmarks | Helen (2330 images) | 256×256 | ▪ Provide more realistic results but sensitive to variation in posture and gaze |
| FSNet [66] | GAN | Facial landmarks | CelebA | 128×128 | ▪ Sensitive to variation in angle |
| RSGAN [65] | GAN | Facial landmarks, segmentation mask | CelebA | 128×128 | ▪ Sensitive to variation in angle, occlusion, lightning<br>▪ Limited output resolution |
| FaceShifter [67] | GAN | Attributes (face, occlusions, lighting or styles) | ▪ VGG Face<br>▪ CelebA-HQ<br>▪ FFHQ | 256×256 | ▪ Stripped artifacts |

**Detection**

As shown in Table 4, attempts were made to detect the faceswap based deepfakes using both handcrafted and deep features.

**Techniques based on handcrafted Features:** Zhang et al. [70] proposed a technique to detect swapped faces by using Speeded Up Robust Features (SURF) descriptor for feature extraction that were then used to train the SVM for classification. This technique was then tested on the set of Gaussian blurred images. This approach has improved the deepfakes image detection performance however, unable to detect manipulated videos. Yang et al. [71] introduced an approach to detect deepfakes by estimating the 3D head position from 2D facial landmarks. The computed difference among the head poses was used as a feature vector to train the SVM classifier and was later used to differentiate between original and forged content. This technique exhibits good performance for deepfake detection but has a limitation in estimating landmark orientation in the blurred images, which degrades the performance of this method under such scenarios. Guera et al. [72] presented a method for detecting synthesized faces from videos. Multimedia stream descriptors [73] were used to extract the features that were then used to train the SVM, and random forest classifiers to differentiate between the real and manipulated faces from the videos. This technique gives an effective solution to deepfakes detection however unable to perform well against video re-encoding attacks. Ciftci et al. [74] introduced an approach to detect forensic changes within videos by computing the biological signals (e.g. heart rate) from the face portion of the videos. Temporal and spatial characteristics of facial features were computed to train the SVM and CNN model to differentiate between bonafide and fake videos. This technique has improved deepfake detection accuracy, however, it has a large feature vector space and its detection accuracy drops significantly when dimensionality reduction techniques are applied. Jung et al. [75] proposed a technique to detect deepfakes by identifying an anomaly based on the time, repetition, and intervened eye-blinking duration within videos. This method combined the Fast-HyperFace [76] and EAR technique (eye detect) [77] to detect eye blinking. An integrity authentication method was employed by tracking the fluctuation of eye blinks based on gender, age, behavior, and time factor to spot the real and fake videos. The approach in [75] exhibits better deepfake detection performance, however, it is not appropriate if the subject in the video is suffering from mental illness as we often experience abnormal eye blinking patterns for such people. Furthermore, the work in [78] [79] have presented ML based approaches for face-swap detection, however, still require performance improvement under the presence of post-processing attacks.

**Techniques based on Deep Features:** Several studies have employed the DL-based method for Face-swap manipulation detection. Li et al. [80] proposed a method of detecting the forensic modifications made within the videos. First, the facial landmarks were extracted using the dlib software package [81]. Next, CNN-based models named ResNet152, ResNet101, ResNet50, and VGG16 were trained to detect forged content from videos. This approach is more robust in detecting the forensic changes; however, it exhibits low performance on multi-time compressed videos. Guera et al. [82] proposed a novel CNN to extract the features at the frame level. Then the RNN was trained on the set of extracted features to detect deepfakes from the input videos. This work achieves good detection performance but only on videos of short duration i.e. videos of 2 seconds or less. Li et al. [83] proposed a technique to detect deepfakes by using the fact that the manipulated videos lack accurate eye blinking in synthesized faces. CNN/RNN approach was used to detect the lack of eye blinking in the videos to expose the forged content. This technique shows better deepfake detection performance, however, it only uses the lack of eye blinking as a clue to detect the deepfakes. This approach has the following potential limitations: i) it is unable to detect the forgeries in videos with frequent eye blinking, ii) it is unable to detect manipulated faces with closed eyes in training, and iii) it is inapplicable in scenarios where forgers can create realistic eye blinking in synthesized faces. Montserrat et al. [84] introduced a method for detecting visual manipulations in a video. Initially, a Multi-task convolutional neural network (MTCNN) [85] was employed to detect the faces from all video frames on which CNN was applied, to compute the features. In the next step, the Automatic Face Weighting (AFW) mechanism, along with a Gated Recurrent Unit, was used to discard the false-detected faces. Finally, an RNN was employed to combine the features from all steps and locate the manipulated content in the videos. The approach in [84] works well for deepfake detection, however, it is unable to obtain the prediction from the features in multiple frames. Lima et al. [86] introduced a technique to detect video manipulation by learning the temporal information of frames. Initially, VGG-11 was employed to compute the features from video frames, on which LSTM was applied for temporal sequence analysis. Several CNN frameworks, named R3D, ResNet, I3D, were trained on the temporal sequence descriptors outputted by the LSTM, to identify original and manipulated videos. This approach [86] improves deepfake detection accuracy but at the expense of high computational cost. Agarwal et al. [87] presented an approach to locate face-swap-based manipulations by combining both facial and behavioral biometrics. The behavioral biometric was recognized with the encoder-decoder network (Facial Attributes-Net, FAb-Net) [88]. Whereas VGG-16 was employed for facial features computation. Finally, by merging both metrics the inconsistencies in the matching identities were revealed to locate face-swap deepfakes. This approach [87] works well for unseen cases, however, it may not generalize well to lip-synch-based deepfakes. Fernandes et al. [89] introduced a technique to locate visual manipulation by measuring the heart-rate of the subjects. Initially, three techniques: skin color variation [90], average optical intensity [91], and Eulerian video magnification [92], were used to measure heart rate. The computed heart-rate was used to train a Neural Ordinary Differential Equations (Neural-ODE) model [93] to differentiate the original and altered content. This technique [89] works well for deepfakes detection but has increased computational complexity. Other works [94-98] have explored CNN based methods for detection of swapped faces, however, still there is a need for more robust approach.

Table 4: An overview of face swap deepfake detection techniques and their limitations

| Author | Technique | Features | Best Evaluation performance | Dataset | Limitations |
|---|---|---|---|---|---|
| **Handcrafted features** | | | | | |
| Zhang et al. [70] | SURF + SVM | 64-D features using SURF | ▪ Precision= 97%<br>▪ Recall= 88%<br>▪ Accuracy= 92% | Generate deepfake dataset using LFW face database. | ▪ Unable to preserve facial expressions<br>▪ Works with static images only. |
| Yang et al. [71] | SVM Classifier | 68-D facial landmarks using DLib | ROC=89%<br>ROC=84% | ▪ UADFV<br>▪ DARPA MediFor GAN Image/ Video Challenge. | ▪ Degraded performance for blurry images. |
| Guera et al. [72] | SVM, RF Classifier | Multimedia stream descriptor [29] | AUC= 93% (SVM)<br>AUC= 96% (RF) | Custom dataset. | ▪ Fails on video re-encoding attacks |
| Ciftci et al. [74] | CNN | medical signals features | Accuracy= 96% | Face Forensics dataset | ▪ Large feature vector space. |
| Jung et al. [75] | Fast-HyperFace[76], EAR[77] | Landmark features | Accuracy= 87.5% | Eye Blinking Prediction dataset | ▪ Inappropriate for people with mental illness |
| Matern et al. [78] | MLP, Logreg | 16-D texture energy based features of eyes and teeth [99] | ▪ AUC= .851(MLP)<br>▪ AUC=0.784 (LogReg) | FF++ | ▪ Only applicable to face images with open eyes and clear teeth. |
| Agarwal et al. [79] | SVM Classifier | 16 AU's using OpenFace2 toolkit | AUC= 93% | Own dataset. | ▪ Degraded performance in cases where a person is looking off-camera. |
| **Deep Learning-based features** | | | | | |
| Li e al. [80] | VGG16, ResNet50, ResNet101, ResNet152 | DLib facial landmarks | AUC=84.5 (VGG16), 97.4 (ResNet50), 95.4 (ResNet101), 93.8 (ResNet152) | DeepFake-TIMIT | ▪ Not robust for multiple video compression. |
| Guera et al. [82] | CNN/ RNN | Deep features | Accuracy=97.1% | Customized dataset | ▪ Applicable to short videos only (2 sec). |
| Li et al. [83] | CNN/RNN | DLib facial landmarks | TPR= 99% | Customized dataset | ▪ Fails over frequent and closed eyes blinking. |
| Montserrat et al. [84] | CNN + RNN | Deep features | Accuracy=92.61% | DFDC | ▪ Performance needs improvement. |
| Lima et al. [86] | VGG11 + LSTM | Deep features | Accuracy= 98.26%, AUC= 99.73% | Celeb-DF | ▪ Computationally complex. |
| Agarwal et al. [87] | VGG6 + encoder-decoder network | Deep features + behavioral biometrics | AUC= 99%<br>AUC= 99%<br>AUC= 93%<br>AUC= 99% | WLDR<br>FF<br>DFD<br>Celeb-DF | ▪ Unable to generalize well to unseen deepfakes. |
| Fernandes et al. [89] | Neural-ODE model | Heart-rate | Loss=0.0215<br>Loss=0.0327 | Custom<br>DeepfakeTIMIT | ▪ Computationally expensive |
| Sabir et al. [94] | CNN/RNN | CNN features | Accuracy= 96.3% | FF++ | ▪ Results are reported for static images only. |
| Afchar et al. [95] | MesoInception-4 | Deep features (DF) | TPR= 81.3 % | FF++ | ▪ Performance degrades on low quality videos. |
| Nguyen et al. [96] | CNN | Deep features | Accuracy=83.71% | FF++ | ▪ Degraded detection performance for unseen cases. |
| Stehouwer et al. [97] | CNN | Deep features | Accuracy=99.43% | Diverse Fake Face Dataset (DFFD) | ▪ Computationally expensive due to large feature vector space. |
| Rossle et al. [98] | SVM + CNN | Co-Occurance matrix + DF | Accuracy= 90.29% | FF++ | ▪ Low performance on compressed videos. |

## 4.2 Lip-syncing
### Generation

The Lip-syncing approach involves synthesizing a video of a target identity such that the mouth region in the manipulated video is consistent with arbitrary audio input [35] (Fig. 5). A key aspect of synthesizing a visual speech is the movement and appearance of the lower portion of the mouth and its surrounding region. To convey a message more effectively and naturally, it is important to generate proper lip movements along with expressions. From a scientific point of view, lip-syncing has many applications in the entertainment industry, such as making audio-driven photorealistic digital characters in films or games, voice-bots, and dubbing films in foreign languages. Moreover, it can also help hearing-impaired persons understand a scenario by lip-reading from a video created using the genuine audio.

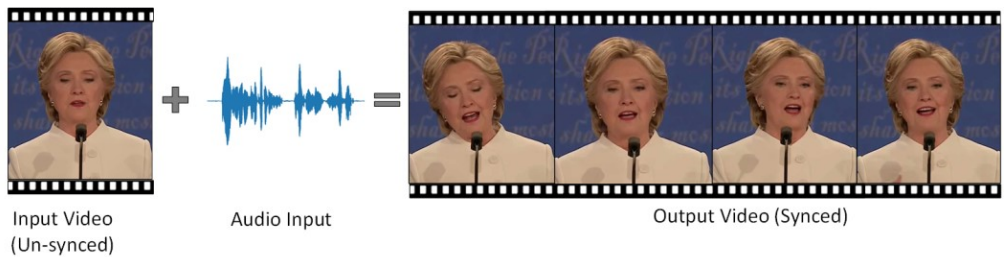

Figure 5: A visual representation of lip-syncing of an existing video to an arbitrary audio clip

Existing works on lip-syncing [100, 101] require the reselection of frames from a video or transcription, along with target emotions, to synthesize the lip's motion. These approaches are limited to a dedicated emotional state or don't generalize well to unseen faces. However, the DL models are capable of learning and predicting the movements from audio features. A detailed analysis of existing methods used for Lip-sync-based deepfakes detection is presented in Table 5. Suwajanakorn et al. [35] proposed an approach to generate a photo-realistic lip-synced video using a target's video and an arbitrary audio clip as input. The recurrent neural network (RNN) based model was employed to learn the mapping between audio features and mouth shape for every frame, and later used frame reselection to fill in the texture around the mouth based on the landmarks. This synthesis was performed on the lower facial regions i.e. mouth, chin, nose, and cheeks. This approach applied a series of post-processing steps, such as smoothing jaw location and re-timing the video to align vocal pauses, or talking head motion, to produce videos that appear more natural and realistic. In this work, Barak Obama was considered as a case study due to the sufficient availability of online video footage. Thus, this model is required to retrain for each individual. The Speech2Vid [102] model took an audio clip and a static image of a target subject as input and generated a video that is lip-synced with the audio clip. This model used the Mel Frequency Cepstral Coefficients (MFCC) features extracted from the audio input and fed them into a CNN-based encoder-decoder. As a post-processing step, a separate CNN was used for frame deblurring and sharpening to preserve the quality of visual content. This model generalizes well to unseen faces and thus does not need retraining for new identities. However, this work is unable to synthesize a variety of emotions on facial expression.

The GAN-based manipulations such as [103] employed a temporal GAN, consisting of an RNN, to generate a photorealistic video directly from a still image and speech signal. The resulting video included synchronized lip movements, eye-blinking, and natural facial expression without relying on manually handcrafted audio-visual features. Multiple discriminators were employed to control frame quality, audio-visual synchronization, and overall video quality. This model can generate lip-syncing for any individual in real-time. In [104], an adversarial learning method was employed to learn the disentangled audio-visual representation. The speech encoder was trained to project both the audio and visual representations into the same latent space. The advantage of using a disentangled representation was that both the audio and video could serve as a source of speech information during the generation process. As a result, it was possible to generate realistic talking face sequences on an arbitrary identity with synchronized lip movement. Garrido et al. [105] presented a Vdub system that captures the high-quality 3D facial model of both the source and the target actor. The computed facial model was used to photo-realistically reconstruct a 3D mouth model of the dubber to be applied on the target actor. An audio channel analysis was performed to better align the synthesized visual content with the audio. This approach better renders a coarse-textured teeth proxy however it fails to synthesize a high-quality interior mouth region. In [106] a face-to-face translation method, LipGAN, was proposed to synthesize a talking face video of any individual utilizing a given single image and audio segment as input. LipGAN consists of a generator network to synthesize portrait video frames with a modified mouth and jaw area from the given audio and target frames and uses a discriminator network to decide whether the synthesized face is synchronized with the given audio. This approach is unable to ensure temporal consistency in the synthesized content, as blurriness and jitter can be observed in the resultant video. Recently, Prajwal et al. [107] proposed a wav2lip speaker-independent model that can accurately synchronize the lips movement in a video recording with a given audio clip. This approach employs a pre-trained lip-sync discriminator that is further trained on noisy generated videos in the absence of a generator. This model uses several consecutive frames instead of a single frame in the discriminator and employs visual quality loss along with contrastive loss, thus increasing the visual quality by considering temporal correlation.

The recent approaches can synthesize photo-realistic fake videos from speech (audio-to-video) or text (text-to-video) with convincing video results. The methods proposed in [35, 108] can edit existing video of a person to the desired speech to be spoken from text input by modifying the mouth movement and speech accordingly. These approaches are more focused on synchronizing lip-movements by synthesizing the region around the mouth. In [109] a VAE based framework was proposed to synthesize full pose video with facial expressions, gestures, and body posture movements from given audio.

Table 5: An overview of Lip sync-based Deepfake generation techniques

| Reference | Technique | Features | Dataset | Output Quality | Limitations |
|---|---|---|---|---|---|
| Suwajanakorn et al. [35] | RNN (single-layer unidirectional LSTM) | ▪ Mouth landmarks (36-D features)<br>▪ MFCC audio features (28-D) | Youtube videos (17 hours) | 2048×1024 | ▪ Requires large amount of training data for target person.<br>▪ Require retraining for each identity.<br>▪ Sensitive to the 3D movement of head<br>▪ No direct control over facial expressions |
| Speech2Vid[102] | Encoder–decoder CNN | ▪ VGG-M network<br>▪ MFCC audio features | ▪ VGG Face<br>▪ LRS2 (41.3-hour video)<br>▪ VoxCeleb2 (test) | 109×109 | ▪ lacks the synthesis of emotional facial expressions |
| Vougioukas et al. [103] | Temporal GAN | MFCC audio features | ▪ GRID<br>▪ TCD TIMIT | 96×128 | ▪ lacks the synthesis of emotional facial expressions<br>▪ flickering and jitter<br>▪ sensitive to large facial motions |
| Zhou et al. [104] | Temporal GAN | Deep audio-video features | ▪ LRW<br>▪ MS-Celeb-1M | 256×256 | ▪ lacks the synthesis of emotional facial expressions |
| Vdub [105] | 3DMM | ▪ 66 facial feature points<br>▪ MFCC features | ▪ Private | 1024×1024 | ▪ Requires video of the target |
| LipGAN [106] | GAN | ▪ VGG-M network<br>▪ MFCC features | ▪ LRS 2 | 1280×720 | ▪ visual artifacts and temporal inconsistency<br>▪ unable to preserve source lip region characteristics |
| Wav2Lip[107] | GAN | ▪ Mel-spectrogram representation | ▪ LRS2 | 1280×720 | ▪ lacks the synthesis of emotional facial expressions |

**Detection**

**Techniques based on handcrafted Features:** Initially, ML-based methods are employed for the detection of lip-sync visual deepfakes. Korshunov et al. [110] proposed a technique employing 40-D MFCC features containing the 13-D MFCC, 13-D delta, and 13-D double-delta, along with the energy, in combination with mouth landmarks to train the four classifiers, i.e. SVM, LSTM, multilayer perceptron (MLP), and Gaussian mixture model (GMM). Three publicly available datasets, named VidTIMIT[111], AMI corpus [112], and GRID corpus [113] were used to evaluate the performance of this technique. From the results, it was concluded in [110] that LSTM achieves better performance over other techniques. However, lip-syncing deepfake detection performance of the LSTM method drops for the VidTIMIT [111] and AMI [112] datasets due to fewer training samples for each person in both of these datasets over the GRID dataset. In [113] MFCC features were substituted with DNN embeddings i.e., language-specific phonetic features used for automatic speaker recognition. The evaluations showed an improved performance as compared to [110], however, performance is not evaluated on large scale realistic dataset and GAN based manipulation.

**Techniques based on Deep Features:** The other DL-based techniques such as [114] proposed a detection approach by exploiting the inconsistencies between phoneme-viseme pairs. In [114] authors observed that in a video the lips shape associated with specific phenomes such as M, B, or P must be completely closed to pronounce them, however, the deepfake videos often lack this aspect. They analyzed the performance by creating deepfakes using Audio-to-Video (A2V) [35] and Text-to-Video (T2V) [108] synthesis techniques. However, it fails to generalize well for unseen samples during training. Haliassos et al. [115] proposed a lip-sync deepfake detection approach namely LipForensics using a spatio-temporal network. Initially, a feature extractor 3D-CNN ResNet18 and a multiscale temporal convolutional network (MS-TCN) are trained on lip-reading dataset such as Lipreading in the Wild (LRW). Then, the model is fine-tuned on deepfake videos using FaceForensics++ (FF++) dataset. The method also performed well over different post-processing operations such as blur, noise, compression etc., however, the performance substantially decreases when there is a limited mouth movement such as pauses in speech or less movement in lips in videos. Chugh et al. [116] proposed a deepfake detection mechanism by finding a lack of synchronization between the audio and visual channels. They computed a modality dissimilarity score (MDS) between the audio and visual modalities. A sub-network based on 3D-ResNet architecture is used for feature computation and employed two loss functions, a cross-entropy loss at the output layer for robust feature learning, and a contrastive loss is computed over segment-level audiovisual features. The MDS is calculated as the total audiovisual dissonance over all segments of the video and is used for the classification of a video as real or fake. Mittal et al. [117] proposed a siamese network architecture for audio-visual deepfake detection. This approach compares the correlation between emotion-based differences in facial movements and speech to distinguish between real and fake. However, this approach requires a real-fake video pair for the training of the network and fails to classify correctly if only a few frames in the video have been manipulated. Chintha et al. [118] proposed a framework based on XceptionNet CNN for facial feature extraction and then passed it to a bidirectional LSTM network for the detection of temporal inconsistencies. The network is trained via two loss functions, i.e. cross-entropy and KL-divergence to discriminate the feature distribution of real video from that of manipulated video. Table 6 presents a comparison of handcrafted and deep learning techniques employed for detection of lip sync-based deepfakes.

Table 6: An overview of Lip sync-based Deepfake detection techniques

| Author | Technique | Performance reported | Dataset used | Limitations |
|---|---|---|---|---|
| **Handcrafted features** | | | | |
| Korshunov et al. [110] | SVM, LSTM, MLP, GMM | EER=24.74 (LSTM), 53.45 (MLP), 56.18(SVM), 56.09(GMM) | VidTIMIT | LSTM performs better than others but its performance degrades as the training samples decrease. |
| | | EER=33.86 (LSTM), 41.21(MLP), 48.39(SVM), 47.84 (GMM) | AMI | |
| | | EER=14.12 (LSTM), 28.58(MLP), 30.06 (SVM), 46.81(GMM) | GRID | |
| Agarwal et al. [114] | SVM | Accuracy=99.6% | Custom dataset | Performance degrades for unseen samples |
| **Deep Learning based features** | | | | |
| Haliassos et al. [115] | 3D-ResNet18, multi-scale temporal convolutional network | AUC=97.1% | FF++ | Performance degrades in cases when there is limited lip movement |
| Mittal et al. [117] | siamese network architecture | Accuracy =84.4% | DFDC | Requires a real–fake video pair for training. |
| | | AUC=96.3%(LQ), 94.9%(HQ) | DF-TIMIT | |
| Chintha et al. [118] | XceptionNet CNN with bidirectional LSTM network | Accuracy=97.83% | Celeb-Df | Performance degrades on compressed samples |
| | | Accuracy=96.89% | FF++ | |

## 4.3 Puppet-master
### Generation
Puppet-master, also known as face reenactment, is another common variation of deepfakes that manipulates the facial expressions of a person e.g., transferring the facial gestures, eye, and head movements to an output video which reflect those of the source actor [119] as shown in Fig. 6. Puppet-mastery aims to deform the person's mouth movement to make fabricated content. Facial reenactment has various applications, i.e. altering the facial expression and mouth movement of a participant to a foreign language in an online multilingual video conference, dubbing or editing an actor's head and their facial expressions in film industry post-production systems, or creating photorealistic animation for movies and games, etc.

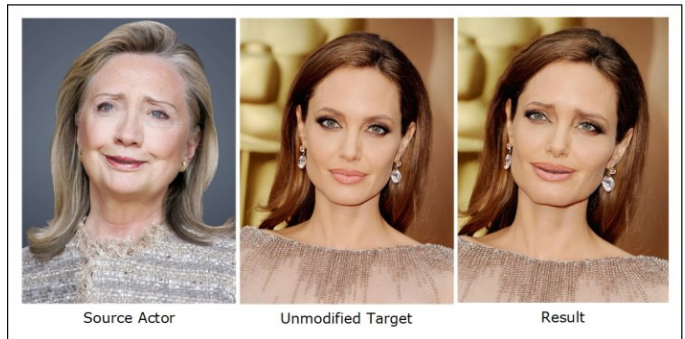
Figure 6: A visual representation of face-reenactment based deepfake

Initially, 3D facial modeling-based approaches for facial reenactment were proposed because of their ability to accurately capture the geometry and movement, and for improved photorealism in reenacted faces. Thies et al. [120, 121] presented the first real-time facial expressions transfer method from an actor to a target person. A commodity RGB-D sensor was used to track and reconstruct the 3D model of a source and target actor. For each frame, the tracked deformations of the source face were applied to the target face model, and later the altered face was blended onto the original target face while preserving the facial appearance of the target face model. Face2Face [36] is an advanced form of facial reenactment technique as presented in [120]. This method works in real-time and is capable of altering the facial movements of generic RGB video streams e.g., YouTube videos, using a standard webcam. The 3D model reconstruction approach was combined with image rendering techniques to generate the output. This creates a convincing and instantaneous re-rendering of the target actor with a relatively simple home setup. This work was further extended to control the facial expressions of a person in a target video based on intuitive hand gestures [122] using an inertial measurement unit [123].

Later, GANs have been successfully applied for facial reenactment due to their ability to generate photo-realistic images. Pix2pixHD [124] produces high-resolution images with better fidelity by combining multi-scale conditional GANs (cGAN) architecture using a perceptual loss. Kim et al. [125] proposed an approach that allows the full reanimation of portrait videos by an actor, such as changing head pose, eye gaze, and blinking, rather than just modifying the facial expression of the target identity and thus produced photorealistic dubbing results. At first, a face reconstruction approach was used to obtain a parametric representation of the face and illumination information from each video frame to produce a synthetic rendering of the target identity. This representation was then fed to a render-to-video translation network based on the cGAN to predict the synthetic rendering into photo-realistic video frames. This approach requires training the videos for target identity. Wu et al. [126] proposed ReenactGAN which encodes the input facial features into a boundary latent space. A target-specific transformer was used to adapt the source boundary space according to the specified target, and later the latent space was decoded onto the target face. GANimation [127] employed a dual cGAN generator conditioned on emotion action units (AU) to transfer facial expressions. The AU-based generator used an attention map to interpolate between the reenacted and original images. Instead of relying on AU estimations, GANnotation [128] used facial landmarks along with the self-attention mechanism for facial reenactment. This approach introduced a triple consistency loss to minimize visual artifacts but requires the images to be synthesized with a frontal facial view for further processing. These models [89-90] require a large amount of training data for target identity to perform well at oblique angles or they will lack the ability to generate photo realistic reenactment for unknown identities.

Recently, few shot or one-shot face reenactment approaches have been proposed to achieve reenactment using a few or even a single source image. In [37], a self-supervised learning model, X2face, using multiple modalities such as driving frame, facial landmarks, or audio to transfer the pose and expression of the input source to target expression, was proposed. X2face used two encoder-decoder networks: an embedding network and a driving network. The embedding network learns face representation from the source frame and the driving network learns pose and expression information from the driving fame to the vector map. The driving network was crafted to interpolate face representation from the embedded network to produce target expressions. Zakharov et al. [129] presented a meta-transfer learning approach where the network was first trained on multiple identities and then fine-tuned on the target identity. First, target identity encoding was obtained by averaging the target's expressions and associated landmarks from different frames. Then a pix2pixHD [124] GAN was used to generate the target identity using source landmarks as input, and identity encoding via adaptive instance normalization (AdaIN) layers. This approach works well at oblique angles and directly transfers the expression without requiring intermediate boundary latent space or interpolation map, as in [37]. Zhang et al. [130] proposed an auto-encoder-based structure to learn the latent representation of the target's facial appearance and source's face shape. These features were used as input to SPADE residual blocks for the face reenactment task, which preserved the spatial information and concatenated the feature map in a multi-scale manner from the face reconstruction decoder. This approach can better handle large pose changes and exaggerated facial actions. In FaR-GAN [131], learnable features from convolution layers were used as input to the SPADE module instead of using multi-scale landmark masks, as in [130]. Usually, few-shot learning fails to completely preserve the source identity in the generated results for cases where there is a large pose difference between the reference and target image. MarioNETte [46] was proposed to mitigate identity leakage by employing attention block and target feature alignment. This helped the model to accommodate the variations between face structures better. Finally, the identity was retained by using a novel landmark transformer, influenced by the 3DMM facial model [132].

The real-time face reenactment approach such as FSGAN [64] performs both facial replacement and reenactment with occlusion handling. For reenactment, a pix2pixHD [124] generator takes the target's image and source's 3D facial landmark as input and outputs a reenacted image and 3-channel (hair, face, and background) encoded segmentation mask. The recurrent generator was trained recursively where output was iterated multiple times for incremental interpolation from source to target landmarks. The results were further improved by applying Delaunay Triangulation and barycentric coordinate interpolation to generate output similar to the target's pose. This method achieves real-time facial reenactment at 30fps and can be applied to any face without requiring identity-specific training. Table 7 provides the summary of techniques adopted for facial expression manipulation and mentioned above.

In the next few years, photo-realistic full-body reenactment [8] videos will also be viable, where the target's expression, along with mannerism, will be manipulated to create realistic deepfakes. The videos that will be generated using the above-mentioned techniques will be further merged with fake audio to create the fabricated content completely [133]. These progressions enable the real-time manipulation of facial expressions and motion in videos while making it challenging to distinguish between real and synthesized video.

Table 7: An overview of face reenactment based Deepfake generation techniques

| Reference | Technique | Features | Dataset | Output Quality | Limitations |
|---|---|---|---|---|---|
| Face2Face [36] | 3DMM | ▪ parametric model ▪ Facial landmark features | customized | 1024×1024 | ▪ Sensitive to facial occlusions |
| Kim et al. [125] | cGAN | parametric model of the face (261 parameters/frame) | customized | 1024×1024 | ▪ 1-3 min. video of target ▪ Sensitive to facial occlusions |
| ReenactGAN [126] | GAN | Facial landmark features | ▪ CelebV dataset ▪ WFLW Dataset ▪ Helen, DISFA | 256×256 | ▪ 30 min. video of target ▪ Lack of gaze adaption |
| GANimation [127] | GAN (2 Encoder- 2 Decoder) | AUs | ▪ EmotioNet dataset ▪ RaFD dataset | 128×128 | ▪ Lack of pose and gaze adaption |
| GANnotation [128] | GAN | Facial landmark features | ▪ 300-VWChallenge dataset ▪ BP4D dataset ▪ Helen, LFPW, AFW, IBUG, and a subset of multiple datasets | 128x128 | ▪ Lack of gaze adaption |
| X2face [37] | 2Encoder-2Decoder | ▪ Facial landmark features ▪ 256-D audio features | ▪ VGG Face dataset ▪ VoxCeleb dataset ▪ AFLW dataset | 256×256 | ▪ Wrinkle artifacts ▪ Lack of gaze adaption |
| Zakharov et al. [129] | GAN (1Encoder-2Decoder) | Facial landmark features | VoxCeleb dataset | 256×256 | ▪ Sensitive to source identity leakage ▪ Lack of gaze adaption |
| Zhang et al. [130] | GAN (1Encoder-2Decoder) | Appearance and shape feature Map | ▪ VGG Face dataset ▪ WFLW ▪ EOTT dataset ▪ CelebA-HQ dataset ▪ LRW dataset. | 256×256 | ▪ Low visual quality output (256×256) |
| FaR-GAN [131] | GAN | Facial landmark and Boundary features | ▪ VGG Face dataset ▪ VoxCeleb1 dataset | 256×256 | ▪ Sensitive to source identity leakage ▪ Lack of gaze adaption |

| | | | | | | |
|---|---|---|---|---|---|---|
| MarioNETte [46] | GAN (2Encoder-1Decoder) | Facial landmark features | ▪ VoxCeleb1 | 256×256 | ▪ Fails to preserve source facial characteristics completely | |
| FSGAN[64] | GAN+RNN | ▪ Facial landmarks<br>▪ LFW parts label set | ▪ IJB-C dataset (5500 face videos)<br>▪ VGGFace2<br>▪ CelebA<br>▪ Figaro dataset | 256×256 | ▪ The identity and texture quality degrade in case of large angular differences<br>▪ Fail to fully capture facial expressions<br>▪ blurriness in image texture<br>▪ limited to the resolution of training data | |

## Detection

**Techniques based on handcrafted Features:** Matern et al. [78] presented an approach for classifying forged content by employing simple facial handcrafted features like the color of eyes, missing artifact information in the eyes and teeth, and missing reflections. These features were used to train two models, i.e. logistic regression and MLP, to distinguish the manipulated content from the original data. This technique has a low computational cost; however, it applies only to the visual content with open eyes or visible teeth. Amerini et al. [134] proposed an approach based on optical flow fields to detect synthesized faces in digital videos. The optical flow fields [135] of each video frame were computed using PWC-Net [136]. The estimated optical flow fields of frames were used to train the VGG16 and ResNet50 to classify bonafide and fake content. This method [134] exhibits better deepfake detection performance, however, only initial results have been reported. Agarwal et al. [79] presented a user-specific technique for deepfakes detection. First, GAN was used to generate all three types of deepfakes for US ex-president Barack Obama. Then the OpenFace2 [137] toolkit was used to estimate facial and head movements. The estimated difference between the 2D and 3D facial and head landmarks was used to train the binary SVM to classify between the original face and synthesized face of Barack Obama. This technique provides good detection accuracy, however, it is vulnerable in those scenarios where a person is looking off-camera.

**Techniques based on Deep Features:** Several research works have focused on employing DL-based methods for puppet-mastery deepfakes detection. Sabir et al. [94] observed that while generating the manipulated content, forgers often do not impose temporal coherence in the synthesis process. So, in [94], a recurrent convolutional model was used to investigate the temporal artifacts to identify synthesized faces in the images. This technique [94] achieves better detection performance, however, it works well on static frames. Rossler et al. [98] employed both the handcrafted (co-occurrence matrix) and learned features for detecting manipulated content. It was concluded in [98] that the detection performance of both networks, either employing hand-crafted or deep features, degrade when evaluating them on compressed videos. To analyze the mesoscopic properties of manipulated content, Afchar et al. [95] proposed an approach where they employed two variants of the CNN model with a small number of layers named Meso-4 and MesoInception-4. This method has managed to reduce the computational cost by downsampling the frames but at the expense of a decrease in accuracy in deepfake detection. Nguyen et al. [96] proposed a multi-task, learning-based CNN network to simultaneously detect and localize manipulated content from the videos. An autoencoder was used for the classification of forged content, while a y-shaped decoder was applied to share the extracted information for the segmentation and reconstruction steps. This model is robust to deepfakes detection; however, the evaluation accuracy degrades over unseen scenarios. To overcome the issue of performance degradation as in [96], Stehouwer et al. [97] proposed a Forensic transfer (FT) based CNN approach for deepfake detection. This work [97], however, suffers from high computational cost due to a large feature space. The comparison of handcrafted and deep features-based face reenactment deepfake detection techniques mentioned above is presented in Table 8.

**Table 8: An overview of face reenactment based Deepfake detection techniques**

| Author | Technique | Features | Best Evaluation performance | Dataset | Limitations |
|---|---|---|---|---|---|
| | | | Handcrafted | | |
| Matern et al. [78] | MLP, Logreg | 16-D texture energy based features of eyes and teeth [99] | ▪ AUC=.823 (MLP)<br>▪ AUC=.866 (LogReg) | FF++ | ▪ Only applicable to face images with open eyes and clear teeth. |
| Agarwal et al. [79] | SVM Classifier | 16 AU's using OpenFace2 toolkit | ▪ AUC= 98% | Own dataset. | ▪ Degraded performance in cases where a person is looking off-camera. |

| Amerini et al. [134] | VGG16, ResNet | Optical flow fields | Accuracy= 81.61% (VGG16), 75.46% (ResNet) | FF++ | ▪ Very few results are reported |
|---|---|---|---|---|---|
| **Deep Learning** | | | | | |
| Sabir et al. [94] | CNN/RNN | CNN features | Accuracy= 94.35 % | FF++ | ▪ Results are reported for static images only. |
| Afchar et al. [95] | MesoInception-4 | Deep features (DF) | TPR= 81.3% | FF++ | ▪ Performance degrades on low quality videos. |
| Nguyen et al. [96] | CNN | Deep features | Accuracy=92.50% | FF++ | ▪ Degraded detection performance for unseen cases. |
| Stehouwer et al. [97] | CNN | Deep features | Accuracy=99.4% | Diverse Fake Face Dataset (DFFD) | ▪ Computationally expensive due to large feature vector space. |
| Rossle et al. [98] | SVM + CNN | Co-Occurance matrix + DF | Accuracy= 86.86% | FF++ | ▪ Low performance on compressed videos. |

## 4.4 Face Synthesis
**Generation**

Facial editing in digital images has been heavily explored for decades. It has been widely adopted in the art, animation, and entertainment industry. However, lately, it has been exploited to create deepfakes for identity impersonation. Face generation involves the synthesis of photorealistic images of a human face that may or may not exist in real life. The tremendous evolution in deep generative models has made them widely adopted tools for face image synthesis and editing. Generative deep learning models, i.e. GAN [139] and VAE [140], have been successfully used to generate photo-realistic fake human face images. In facial synthesis, the objective is to generate non-existent but realistic-looking faces. Face synthesis has enabled a wide range of beneficial applications, like automatic character creation for video games and 3D face modeling industries. AI-based face synthesis could also be used for malicious purposes, as the synthesis of photorealistic fake images for social network accounts identity to spread misinformation. Several approaches have been proposed to generate realistic-looking facial images that humans are unable to recognize as to whether they are real or synthesized. Fig. 7 shows synthetic facial images and the improvement in their quality between 2014 and 2019 that are nearly indistinguishable from real photographs. Table 9 provides a summary of works presented for generation entire synthetic faces.

Since the emergence of GAN [139] in 2014, significant efforts have been made to improve the quality of synthesized images. The images generated using the first GAN model [139] were low-resolution and not very convincing. DCGAN [141] was the first approach that introduced a deconvolution layer in the generator to replace the fully connected layer, which achieved better performance in synthetic image generation. Liu et al. [142] proposed CoGAN, based on VAE, for learning joint distributions of two-domain images. This model trained a couple of GANs rather than a single one, and each was responsible for synthesizing images in one domain. The size of generated images still remained relatively small, e.g. 64×64 or 128×128 pixels.

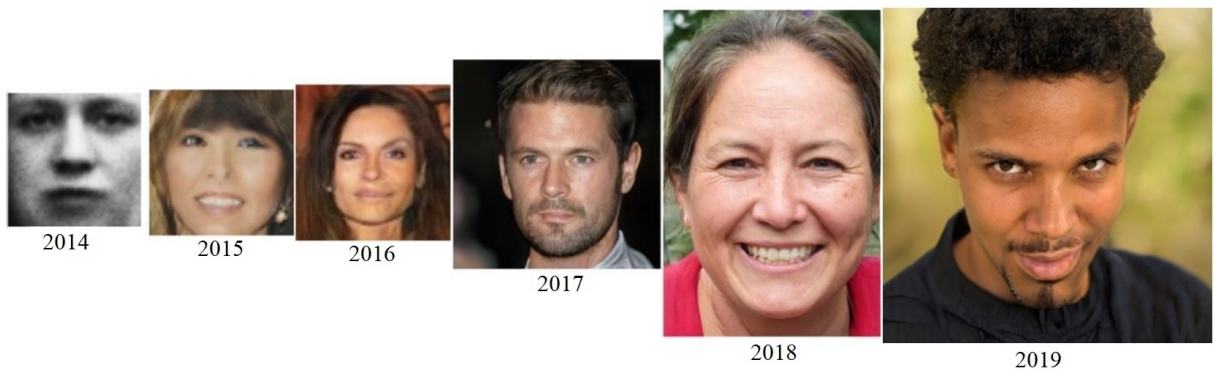

**Figure 7: Increasingly improving improvements in the quality of synthetic faces, as generated by variations on GANs. In order, the images are from papers by Goodfellow et al. (2014) [139], Radford et al. (2015) [141], Liu et al. (2016) [142], Karras et al. (2017) [143], and Style-based (2018 [144], 2019 [145])**

The generation of high-resolution images was limited earlier due to memory constraints. Karras et al. [143] presented ProGAN, a training methodology for GANs, that employed an adaptive mini-batch size that progressively increased the resolution, depending on the current output resolution, by adding layers to the networks during the training process. StyleGAN [144] is an improved version of ProGAN [143]. Instead of mapping latent code z to a resolution, a Mapping Network was employed that learned to map input latent vector (Z) to an intermediate latent vector (W) which controlled different visual features. The improvement is that the intermediate latent vector is free from any certain distribution restriction, and this reduces the correlation between features (disentanglement). The layers of the generator network are controlled via an AdaIN operation which helps decide the features in the output layer. Compared to [139, 141, 142], StyleGAN [144] achieved state-of-the-art high resolution in the generated images i.e., 1024 × 1024, with fine details. StyleGAN2 [145] further improved the perceived image quality by removing unwanted artifacts, such as a change in gaze direction and teeth alignment, with the facial pose. Huang et al. [146] presented a Two-Pathway Generative Adversarial Network (TP-GAN) that could simultaneously perceive global structures and local details, like humans, and synthesize a high-resolution frontal view facial image from a single ill-posed face image. Image synthesis using this approach preserves the identity under large pose variations and illumination. Zhang et al. [147] introduced a self-attention module in convolutional GANs (SAGAN) to handle global dependencies, and thus ensured that the discriminator can accurately determine the related features in distant regions of the image. This work further improved the semantic quality of the generated image. In [148], the authors proposed BigGAN architecture, which uses residual networks to improve image fidelity and the variety of generated samples by increasing the batch size and varying latent distribution. In BigGAN, the latent distribution is embedded in multiple layers of the generator to influence features at different resolutions and levels of the hierarchy rather than just adding to the initial layer. Thus, the generated images were photo-realistic and very close to real-world images from the ImageNet dataset. Zhang et al. [149] proposed a stacked GAN (StackGAN) model to generate high-resolution images (e.g., 256×256) with details based on a given textual description.

Table 9: An overview of synthetic facial deepfake generation techniques

| Reference | Technique | Features | Dataset | Output Quality | Limitations |
|---|---|---|---|---|---|
| Liu et al. [142] | CoGAN | Deep Features | CelebA | 64×64 or 128×128 | ▪ Generate low-quality samples |
| Karras et al. [143] | ProGAN | Deep Features | CelebA | 1024×1024 | ▪ Limited control on the generated output |
| Karras et al. [145] | StyleGAN | Deep Features | ▪ ImageNet | 1024×1024 | ▪ Blob-like artifacts |
| Huang et al. [146] | TP-GAN | Deep Features | ▪ LFW | 256x256 | ▪ Lack fine details<br>▪ Lack semantic consistency |
| Zhang et al. [147] | SAGAN | Deep Features | ▪ ImageNet2012 | 128×128 | ▪ Unwanted visible artifacts |
| Brock et al. [148] | BigGAN | Deep Features | ▪ ImageNet | 512×512 | ▪ Class-conditional image synthesis<br>▪ Class leakage |
| Zhang et al. [149] | StackGAN | Deep Features | ▪ CUB<br>▪ Oxford<br>▪ MS-COCO | 256×256 | ▪ Lack semantic consistency |

**Detection**
**Techniques based on handcrafted Features:** A lot of literature is available on image forgery detection [150-153]. As AI-manipulated data is a new phenomenon, there are a small number of forensic techniques that work well for deepfake detection. Recently, some researchers [70, 154] have adopted the idea of employing the traditional methods of image forgery identification to detect synthesized faces, however, these approaches are unable to identify fake facial images**.** Currently, researchers have focused on new ML-based techniques such as McCloskey et al. [155] presented an approach to identify fake images by employing the fact that the color information is dissimilar between the real camera and fake synthesis samples. The color key-points from input samples were used to train the SVM for classification. This approach [155] exhibits better fake sample detection accuracy, however, it may not perform well for blurred images. Guarnera et al. [156] proposed a method to identify fake images. Initially, the EM algorithm was used to calculate the image features. The computed key-points were used to train three types of classifiers, KNN, SVM, and LDA. The approach in [156] performs well for synthesized image identification, but may not perform well for compressed images.
**Techniques based on Deep Features:** DL-based work such as Guarnera et al. [156] presented an approach to detect image manipulation. Initially, Expectation-Maximization (EM) technique was applied to obtain the image features based on which the naive classifier was trained to discriminate against original and fake images. This approach shows better deepfake identification accuracy, however, it is only applicable to static images. Nataraj et al. [138] proposed a method to detect forged images by calculating the pixel co-occurrence matrices at three color channels of the image. Then a CNN model was trained to learn important features from the co-occurrence matrices to differentiate manipulated and non-manipulated content. Yu et al. [157] presented an attribution network architecture to map an input sample to its related fingerprint image. The correlation index among each sample fingerprint and model fingerprint acts as a softmax logit for classification. This approach [157] exhibits better detection accuracy, however, it may not perform well with post-processing operations i.e. noise, compression, and blurring, etc. Marra et al. [158] proposed a study to identify the GAN-generated fake images. Particularly, [158] introduced a multi-task incremental learning detection approach to locate and classify new types of GAN-generated samples without affecting the detection accuracy of the previous ones. Two solutions related to the position of the classifier were introduced by employing the iCaRL algorithm for incremental learning [159], named as Multi-Task MultiClassifier, and Multi-Task Single Classifier. This approach [158] is robust to unseen GAN-generated samples but unable to perform well if the information on the fake content generation method is not available. Table 10 presents the comparison of face synthesis deepfake detection techniques mentioned above.

Table 10: An overview of synthetic facial deepfake detection techniques

| Author | Technique | Features | Best Evaluation performance | Dataset | Limitations |
|---|---|---|---|---|---|
| **Handcrafted** | | | | | |
| Guarnera et al. [156] | EM + (KNN, SVM, LDA) | Deep features | ▪ Accuracy=99.22 (KNN)<br>▪ Accuracy= 99.81(SVM)<br>▪ Accuracy= 99.61 (LDA) | CelebA | Not robust to compressed images. |
| McCloskey et al. [155] | SVM | Color channels | ▪ AUC=70% | MFC2018 | Performance degrades over blurry samples. |
| **Deep Learning** | | | | | |
| Nataraj et al [138] | CNN | Deep features + co-occurrence matrices | Accuracy = 99.49% | ▪ cycleGAN | ▪ Works with static images only.<br>▪ Low performance for jpeg compressed images. |
| | | | Accuracy = 93.42% | ▪ StarGAN | |
| Yu et al. [157] | CNN | Deep features | Accuracy = 99.43% | CelebA | ▪ Poor performance on post-processing operations. |
| Marra et al. [158] | CNN + Incremental Learning | Deep features | Accuracy = 99.3% | Customized | ▪ Needs source manipulation technique information |

## 4.5 Facial Attribute Manipulation
**Generation**
Face attribute editing involves altering the facial appearance of an existing sample by modifying the attribute-specific region while keeping the irrelevant regions unchanged. Face attribute editing includes removing/wearing eyeglasses, changing viewpoint, skin retouching (e.g., smoothing skin, removing scars, and minimizing wrinkles), and even some higher-level modifications, such as age and gender, etc. Increasingly, people have been using commercially available AI-based face editing and mobile applications such as FaceApp [3] to automatically alter the appearance of an input image.

Recently, several GAN-based approaches have been proposed to edit facial attributes, such as the color of the skin, hairstyle, age, and gender by adding/removing glasses and facial expression, etc., of the given face. In this manipulation, the GAN takes the original face image as input and generates the edited face image with the given attribute, as shown in Fig. 8. A summary of face attribute manipulation approaches is presented in Table 11. Perarnau et al. [160] introduced the Invertible Conditional GAN (IcGAN), which uses an encoder in combination with cGANs for face attribute editing. The encoder maps the input face image into latent representation and attributes manipulation vector and cGAN reconstructs the face image with new attributes given the altered attributes vector as the condition. This suffers from information loss and alters the original face identity in the synthesized image. In [161], a Fader Network was presented, where an encoder-decoder architecture was trained in an end-to-end manner which generated an image by disentangling the salient information of the image and the attribute values directly in latent space. This approach, however, adds unexpected distortion and blurriness, and thus fails to preserve the original fine details in the generated image.

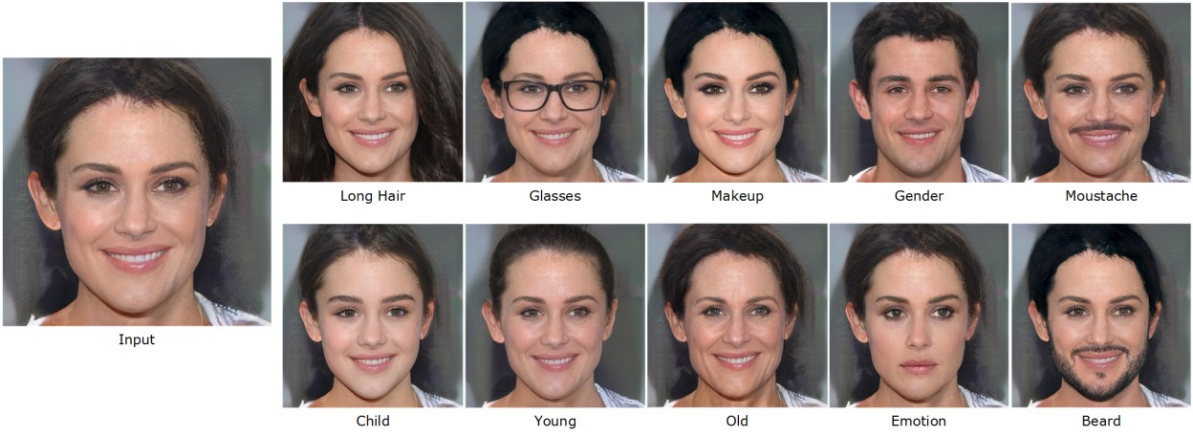

**Figure 8: Examples of different face manipulations: original sample (Input) and manipulated samples**

Prior studies [160, 161] have been focused on handling image-to-image translations between two domains. These methods required the different generators to be trained independently to handle translations between each pair of image domains and thus limit their practical usage. StarGAN [34], an enhanced approach, is capable of translating images among multiple domains using a single generator. A conditional facial attribute transfer network was trained via attribute classification loss and cycle consistency loss. StarGAN achieved promising visual results in terms of attribute manipulation and expression synthesis. However, this approach adds some undesired visible artifacts in the facial skin such as the uneven color tone in the output image. The recently proposed StarGAN-v2 [162] achieved state-of-the-art visual quality of the generated images as compared to [34] by adding a random Gaussian noise vector into the generator. In AttGAN [163], an encoder-decoder architecture was proposed that considers the relationship between attributes and latent representation. Instead of imposing an attribute independent constraint on latent representation like in [160, 161], an attribute classification constraint was applied to the generated image to guarantee the correct change of the desired attributes. AttGAN provided improved facial attribute editing results, with other facial details well preserved. However, the bottleneck layer i.e., down-sampling in the encoder-decoder architecture, adds unwanted changes and blurriness and generates low-quality edited results. Liu et al. [164] proposed the STGAN model that incorporated an attribute difference indicator and a selective transfer unit with an encoder-decoder to adaptively select and modify the encoded features. STGAN only focuses on the attribute-specific region and does not guarantee good preservation of the details in attribute-irrelevant regions.

Other works introduce the attention mechanism for attribute manipulation. SAGAN [165] introduced a GAN-based attribute manipulation network to perform alteration and a global spatial attention mechanism to localize and explicitly constrain editing within a specified region. This approach preserves the irrelevant details well but at the cost of attribute correctness in the case of multiple attribute manipulation. PA-GAN [166] employed a progressive attention mechanism in GAN to progressively blend the attribute features into the encoder features constrained inside a proper attribute area by employing an attention mask from high to low feature level. As the feature level gets lower (higher resolution), the attention mask gets more precise and the attribute editing becomes fine. This approach successfully performs the multiple attributes manipulation and well preserves irrelevance within a single model. However, some undesired artifacts appear in cases where significant modifications are required such as baldness and open mouth.

**Table 11: An overview of facial attribute manipulation based Deepfake generation techniques**

| Author | Technique | Features | Best Evaluation performance | Dataset | Limitations |
|---|---|---|---|---|---|
| Perarnau et al. [160] | IcGAN | ▪ Deep Features | ▪ CelebA<br>▪ MNIST | 64×64 | ▪ Fails to preserve original face identity |
| Fader Network [161] | Encoder-decoder | ▪ Deep Features | ▪ CelebA | 256×256 | ▪ Unwanted distortion and blurriness<br>▪ Fails to preserve fine details |
| Choi et al. [162] | StarGAN | ▪ Deep Features | ▪ CelebA<br>▪ RaFD | 512×512 | ▪ Undesired visible artifacts in the facial skin e.g., the uneven color tone |
| He et al. [163] | AttGAN | ▪ Deep Features | ▪ CelebA<br>▪ LFW | 384 × 384 | ▪ Generates low-quality results and adds unwanted changes, blurriness |
| Liu et al. [164] | STGAN | ▪ Deep Features | ▪ CelebA | 384×384 | ▪ Poor performance for multiple attribute manipulation |

| Zhang et al. [165] | SAGAN | ▪ Deep Features | ▪ CelebA | 256×256 | ▪ Lack of details in the attribute-irrelevant region |
| He et al. [166] | PA-GAN | ▪ Deep Features | ▪ CelebA | 256×256 | ▪ undesired artifacts in case of baldness and open mouth etc. |

**Detection**

**Techniques based on handcrafted Features:** Researchers have employed the traditional ML-based approaches for the detection of facial attributes manipulation. Like in [167], the author used the pixel co-occurrence matrices to compute the features from the suspected samples. The extracted keypoints were used to train a CNN classifier to differentiate the original and manipulated faces. The method in [167] shows better facial attribute manipulation detection accuracy, however, may not perform well over the noisy samples. An identification approach using keypoints computed from the frequency domain, instead of employing raw sample pixels, was introduced in [168]. For each input sample, a 2D DFT was applied to transform the image to the frequency domain to acquire one frequency sample per RGB channel. For predicting the real and fake samples, the work [168] used the AutoGAN classifier. The generalization ability of the work in [168] was evaluated over unseen GAN frameworks. More specifically, they considered two GAN frameworks namely StarGAN [34] and the GauGAN [169]. The work shows better prediction accuracy for the StarGAN model, however, in the case of GauGAN the technique faces serious performance drop.

**Techniques based on Deep Features:** The research community has presented several methods to detect facial manipulations by evaluating the internal GAN pipeline. Similar work was presented in [170] where the author gave the concept that analyzing the internal neuron behaviors could assist in identifying the manipulated faces, as layer-by-layer neuron activation arrangements extract a more representative set of image features which are significant for recognizing the original and fake faces. The proposed solution in [170] namely FakeSpoter, computed the deep features via employing several DL-based face recognition frameworks i.e. VGG-Face [171], OpenFace [172], and FaceNet [173]. The extracted features were used to train the SVM classifier to categorize the fake and real faces. The work [170] worked well for facial attributes manipulation detection, however, it may not perform well for samples with intense light variations.

Existing works on facial attribute manipulation have either employed entire faces or pass face patches to spot real and manipulated content. A face patch-based technique was presented in [174], where the Restricted Boltzmann Machine (RBM) was used to compute the deep features. Then, the extracted features were used to train a two-class SVM classifier to classify the real and forged faces. The method in [174] is robust to manipulated face detection, however, at the expense of increased computational cost. Another similar approach was proposed in [175], where a CNN-based keypoints extractor was presented. The CNN approach comprised 6 convolutional layers along with 2 fully connected layers. Additionally, residual connections were introduced encouraged from the ResNet frameworks to compute the deep features from the input samples. Finally, the calculated features were used to train the SVM classifier to predict the real and manipulated faces. The approach in [175] shows better manipulation identification performance, however, does not report the results over various post-processing attacks i.e. noise, blurring, intensity variations, and color changes. Some researchers have employed the entire faces instead of using the face patches to detect the facial attribute manipulation from visual content. One of such works was presented by Tariq et al. [176] where several DL-based frameworks i.e. VGG-16, VGG-19, ResNet, and XceptionNet were trained over the suspected samples to locate the digital facial attribute forgeries. The work in [176] shows better face attribute manipulation detection performance, however, its performance degrades for real-world scenarios. Some works use attention mechanisms to further enhance the training procedure of the attribute manipulation detection systems. Dang et al. [177] introduced a framework to locate several types of facial manipulations. This work employed attention mechanisms to enhance the feature maps calculation procedures of CNN frameworks. In the case of face attribute manipulation recognition, two different methods of attribute manipulation generation were taken: i) fake samples generated by using the public FaceApp software, by considering various available filters ii) fake samples generated with the StarGAN network. The work [177] is robust to face forgeries detection, however, at the expense of enhanced economic burden.

Wang et al. [164] proposed a framework to detect the manipulated faces. The proposed solution comprised two classification steps namely local and global predictors. For global estimation, a new model namely Dilated Residual Networks (DRN) was used to predict the real and fake samples. While for local estimation, the optical flow fields were utilized. The approach [164] works well for face attribute manipulation identification, however, requires extensive training data. Similarly, the work in [158] proposed a DL-based framework namely XceptionNet for the face attributes forgeries detection and show robust performance. However, the method in [158] is suffering from high computational costs. Rathgeb et al. [178] introduced a face attribute manipulation recognition framework namely Photo Response Non-Uniformity (PRNU). More precisely, scores gathered after performing the analysis of spatial and spectral features computed from the PRNU patterns from entire image samples were fused. The approach [178] is robust to differentiate between the bonafide and retouched facial samples, however, detection accuracy needs further improvements

To conclude the face attribute manipulation detection section, we can say that most of the existing detection work is based on employing DL-based approaches and showing robust performance close to 100% as shown in Table 12. The main reason for the accurate detection accuracy of models is due to the presence of GAN fingerprint information in the manipulated samples. However, now, the researchers have presented such approaches which have removed the fingerprints from the forged samples while maintaining the image realism which is showing a new challenge even for the high-performing attribute manipulation detection frameworks.

Table 12: An overview of facial attribute manipulation based deepfake detection techniques

| Author | Technique | Features | Best Evaluation performance | Dataset | Limitations |
|---|---|---|---|---|---|
| Hand-crafted | | | | | |
| [167] | co-occurrence matrices along with CNN | Co-Occurrence matrix | Accuracy = 99.4% | Private dataset. | ▪ Its evaluation performance reduces over noisy images. |
| [168] | GAN Discriminator | Frequency domain features | Accuracy =100% | Private dataset. | ▪ The technique faces serious performance degradation for GauGAN framework-based face attribute manipulations. |
| Deep Learning | | | | | |
| [170] | FakeSpoter | Deep features | Accuracy = 84.7% | Private dataset | ▪ Its detection performance degrades over the samples with intense light variations. |
| [174] | RBM along with the SVM classifier | Deep features | Accuracy = 96.2% | Private dataset | ▪ This method is suffering from the high computational cost. |
| | | | Accuracy = 87.1% | Private dataset (Celebrity Retouching, ND-IIITD Retouching) | |
| [175] | CNN + SVM | Deep features | Accuracy =99.7% | Private dataset | ▪ Results are reported for post-processing attacks. |
| [176] | CNNs | Deep features | AUC=74.9% | Private dataset | ▪ Performance degrades for real-world scenarios. |
| [177] | Attention Mechanism along with CNN | Deep features | AUC=99.9% | DFFD | ▪ This work is computationally complex. |
| [164] | DRN | Deep features | Average precision=99.8% | Private dataset | ▪ The approach should be evaluated over a standard dataset. |
| [158] | Incremental Learning along with the CNN | Deep features | Accuracy =99.3% | Private dataset | ▪ This work is economically inefficient. |
| [178] | Score-Level Fusion | PRNU Features | EER=13.7% | Private dataset | ▪ The work needs to improve the classification accuracy. |

## 4.6 Audio Deepfakes Generation

AI-synthesized audio manipulation is a type of deepfake that can clone a person's voice and depict that voice saying something outrageous, that the person never said. Recent advancements in AI-synthesized algorithms for speech synthesis and voice cloning have shown the potential to produce realistic fake voices that are nearly indistinguishable from genuine speech. These algorithms can generate synthetic speech that sounds like the target speaker based on text or utterances of the target speaker, with highly convincing results [55, 179]. The synthetic voice is widely adapted for the development of different applications, such as automated dubbing for TV and film, chatbots, AI assistants, text readers, and personalized synthetic voices for vocally handicapped people. Aside from this, synthetic/fake voices have become an increased threat to voice biometric systems [180] and can be used for malicious purposes, such as political gains, fake news, and fraudulent scams, etc. More complex audio synthesis could be combining the power of AI and manual editing. For example, neural network-powered voice synthesis models, such as Google's Tacotron [53],

Wavenet [52] or AdobeVoco [181], can generate realistic and convincing fake voices that resemble the victim's voice, as the first step. Later on, audio editing software, e.g. Audacity [4], can be used to combine the different pieces of original and synthesized audios to make more powerful audios.

AI-based impersonation is not limited to visual content; recent advancements in AI-synthesized fake voices are assisting the creation of highly realistic deepfakes videos [35]. These developments in speech synthesis have shown their potential to produce realistic and natural audio deepfakes, exhibiting real threats to society [182]. Combining synthetic audio content with visual manipulation can significantly make deepfake videos more convincing and increase their harmful impact [35]. Despite much progress, these synthesized speeches lack some aspects of voice quality, like expressiveness, roughness, breathiness, stress, and emotion, etc., specific to a target identity [183]. The AI research community is doing efforts to overcome these challenges and produce human-like voice quality with high speaker similarity.

The different modalities for audio deepfakes are TTS synthesis and VC. TTS synthesis is a technology that can synthesize the natural-sounding voice of any speaker based on the given input text [184]. VC is a technique that modifies the audio waveform of a source speaker to a sound similar to the target speaker's voice [185]. A VC system takes an audio-recorded file of an individual as a source and creates a deepfake audio of the target individual. It preserves the linguistic and phonetic characteristics of the source utterance and emphasis the naturalness and similarity to that of the target speaker. TTS synthesis and VC represent a genuine threat as both generate completely synthetic computer-generated voices that are nearly indistinguishable from genuine speech. Moreover, the cloned replay attacks [12] impose a potential risk for voice biometric devices because the latest speech synthesis techniques can produce voice with high speaker similarity [186]. This section lists the latest progress in speech synthesis including TTS and voice conversion techniques as well as detection techniques.

**TTS Voice Synthesis**

TTS is a decades-old technology that can synthesize the natural-sounding voice of a speaker from a given input text, and thus enables a voice to be used for better human-computer interaction. The initial research on TTS synthesis technology has been done using the methods of speech concatenation or parameter estimation. The concatenative TTS systems are based on separating high-quality recorded speech into small fragments followed by concatenation into a new speech. In recent years, this method has become outdated and unpopular as it is not scalable and consistent. In contrast, parametric models map the text to the salient parameters of the speech, and convert them into an audio signal using the vocoders. Later on, the deployment of deep neural networks gradually become a dominant method for speech synthesis that achieved much better voice quality. These methods include Neural vocoders [57-62], GAN[63-64], autoencoder [65], autoregressive models [52, 53, 187], and other emerging techniques [188-192] have promoted the rapid development of the speech synthesis industry. Fig. 9 shows the principle design of modern TTS methods.

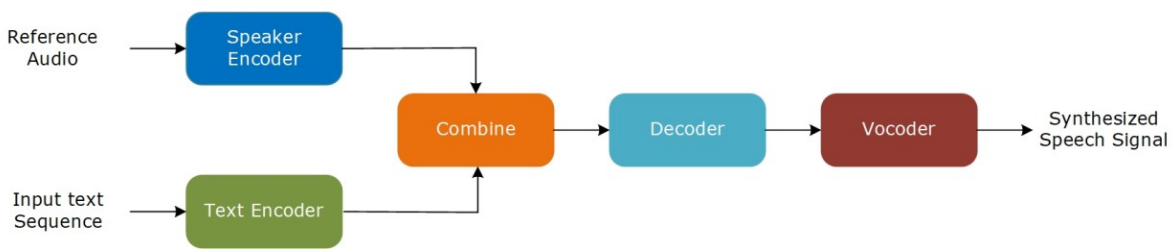

**Figure 9: Workflow diagram of the latest TTS systems**

The significant developments in voice/speech synthesis are WaveNet [52], Tacotron [53], and DeepVoice3 [193], which can generate realistic sounding synthetic speech from a text input to provide an enhanced interaction experience between humans and machines. Table 13 presents an overview of state-of-the-art speech synthesis methods. WaveNet [52] was developed by DeepMind in 2016 and evolved from pixelCNN [194]. WaveNet models utilize raw audio waveforms by using acoustic features, i.e. spectrograms, through a generative framework that is trained on actual recorded speech. Parallel WaveNet has been introduced to enhance the sampling efficacy and produce high-fidelity audio signals [195]. Another DL based using a variant of WaveNet, namely Deep Voice 1 [54], is presented by replacing each module containing an audio signal, voice generator, or a text analysis front-end through a related NN model. Due to the independent training of each module, however, it is not a real end-to-end speech synthesis system. In 2017, Google introduced tacotron [53] an end-to-end speech synthesis model. Tacotron can synthesize speech from given <text, audio> pairs and thus generalizes well to other datasets. Similar to WaveNet, the Tacotron framework is a generative framework comprised of a seq2seq model that contains an encoder, an attention-based decoder, and a post-processing network. Even though the Tacotron model has attained better performance it has one potential limitation i.e. it must employ multiple recurrent components. The inclusion of these units makes it economically inefficient so that it requires high-performance systems for model training. Deep Voice 2 [196] combines the capabilities of both the Tacotron and WaveNet models for voice synthesis. Initially, Tacotron is employed for converting the input text to a linear scale spectrogram, then it is later converted to voice through the WaveNet model. In [197], Tacotron2 was introduced for vocal synthesis and it exhibits an impressive high mean opinion score very similar to human speech. Tacotron2 consists of a recurrent sequence-to-sequence keypoint estimation framework that maps character embedding to mel-scale spectrograms. To deal with the time complexities of recurrent unit-based speech synthesis models, a new, fully-convolutional character-to-spectrogram model named DeepVoice3 [193] was presented. The Deep Voice 3 model is faster than its peers due to performing fully parallel computations. Deep Voice 3 is comprised of three main modules: i) an encoder that accepts text as input and transforms it into an internal learned form, ii) a decoder that converts the learned representations in an autoregressive manner, and iii) post-processing, fully convolutional network that predicts the final vocoder parameters.

Another model for voice synthesis is VoiceLoop [187], which uses a memory framework to generate speech from voices unseen during training. VoiceLoop builds a phonological store by executing a shifting buffer as a matrix. Text strings are characterized as a list of phonemes that are later decoded in short vectors. The new context vector is produced by assessing the encoding of the resulting phonemes and summing them together. The above-mentioned powerful end-to-end speech synthesizer models [193, 197] have enabled the production of large-scale commercial products, such as Google Cloud TTS, Amazon AWS Polly, and Baidu TTS. All these projects aim to attain a high similarity between synthesized and human voices.

The latest TTS systems can convert given text to a human speech with a particular voice identity. Using generative models, researchers have built voice imitating TTS models that can clone the voice of a particular speaker in real-time using few samples of reference speech samples [188, 189]. The key distinction between voice cloning and speech synthesis systems is that the former focuses on preserving the characteristics of the specific identity speech attributes while the latter lacks this feature to maintain the quality of the generated speech [190]. Various AI-enabled voice cloning online platforms are available such as Overdub[1], VoiceApp[2], and iSpeech[3] which can produce synthesized fake voices that closely resemble target speech and gives the public access to this technology. Jia et al. [188] proposed a Tacotron 2 based TTS system capable of producing multi-speaker speech, including those unseen during training. The framework consists of three independently trained neural networks. The findings show that although the synthetic speech resembles a target speaker's voice it does not fully isolate the voice of the speaker from the prosody of the audio reference. Arik et al. [55] proposed a Deep Voice 3 based technique comprised of two modules: speaker adaptation and speaker encoding. For speaker adaptation, a multi-speaker generative framework is fine-tuned. For speaker encoding, an independent model is trained to directly infer a new speaker embedding, which is applied to the multi-speaker generative model.

Loung et al. [190] proposed a voice cloning framework that can synthesize target-specific voice, either from input text or a reference raw audio waveform from a source speaker. The framework consists of a separate encoder and decoder for text and speech and a neural vocoder. The model is jointly trained with linguistic latent features and the speech generation model learns a speaker-disentangled representation. The obtained results achieve quality and speaker similarity to the target speaker; however, it takes almost 5 minutes to generate the cloned speech. Chen et al. [191] proposed a meta-learning approach using waveNet model for voice adaption with limited data. Initially, speaker adaptation is computed by fine-tuning the speaker embedding. Then a text-independent parametric approach is applied whereby an auxiliary encoder network is trained to predict the embedding vector of new speakers. This approach performs well on clean and high-quality training data. The presence of noise deviates the speaker encoding and directly affects the performance of synthesized speech. In [192], the authors proposed a seq2seq multi-speaker framework with domain adversarial training to produce a target speaker voice from only a few available noisy samples. The results showed improved naturalness of synthetic speech. However, similarity still remains challenging to achieve due to lack of transferring target accents, and prosody to synthesized speech with a limited amount of low-quality speech data.

Table 13: An overview of the state-of-the-art speech synthesis techniques

| Methods | Technique | Features | Dataset | Limitations |
| --- | --- | --- | --- | --- |

---

[1] https://www.descript.com/overdub
[2] https://apps.apple.com/us/app/voiceapp/id1122985291
[3] https://www.ispeech.org/apps

| Method | Architecture | Features | Dataset | Limitations |
|---|---|---|---|---|
| WaveNet [52] | Deep neural network | - linguistic features<br>- fundamental frequency (log F0) | - VCTK (44 hrs.) | - Computationally expensive |
| Tacotron [53] | Encoder-Decoder with RNN | - Deep features | Private (24.6 hrs.) | - Costly to train the model |
| Deep Voice 1[54] | Deep neural networks | - linguistic features | Private (20 hrs.) | - Independent training of each module leads to a cumulative error in synthesized speech |
| Deep Voice 2 [196] | RNN | - Deep features | VCTK (44 hrs.) | - Costly to train the model |
| DeepVoice3 [193] | Encoder-decoder | - Deep features | - Private (20 hrs.)<br>- VCTK (44 hrs.)<br>- LibriSpeech ASR (820 hrs.) | - Does not generalized well for unseen samples. |
| Parallel WaveNet [195] | Feed-forward neural network with dilated causal convolutions | - linguistic features | Private | - Requires a large amount of target's speech training data. |
| VoiceLoop [187] | Fully-connected neural network | - 63-dimensional audio features | - VCTK (44 hrs.)<br>- Private | - Low ecological validity |
| Tacotron2[197] | - Encoder-decoder | - linguistic features | - Japanese speech corpus from the ATR Ximera dataset (46.9 hrs.) | - Lack of real time speech synthesis |
| Arik et al. [55] | Encoder-decoder | - Mel spectrograms | - LibriSpeech (820 hrs.)<br>- VCTK (44 hrs.) | - Low performance for multi-speaker speech generation in the case of low-quality audio |
| Jia et al. [188] | Encoder-decoder | - Mel spectrograms | - LibriSpeech (436 hrs.)<br>- VCTK (44 hrs.) | - Fails to attain human-level naturalness<br>- Lacks in transferring the target accent, prosody to synthesized speech |
| Luong et al. [190] | Encoder-decoder | - Mel spectrograms | - LibriSpeech (245 hrs.)<br>- VCTK (44 hrs.) | - Low performance in the case of noisy audio samples |
| Chen et al. [191] | Encoder + deep neural network | - Mel spectrograms | - LibriSpeech (820 hrs.)<br>- private | - Low performance in the case of a low-quality audio sample |
| Cong et al. [192] | Encoder-decoder | - Mel spectrograms | - MULTI-SPK<br>- CHiME-4 | - Lacks in synthesizing utterances of a target speaker |

**Voice Conversion**

VC is speech-to-speech synthesis technology that manipulates the input voice to sound like target voice identity while the linguistic content of the source speech remains unchanged. VC has various applications in real life including expressive voice synthesis, personalized speech speaking assistance, vocally impaired people, voice dubbing for the entertainment industry, and many others [185]. The recent development of anti-spoofing for automated speaker verification [180] included VC systems for the generation of spoofing data [198-200].

In general, to perform VC, high-level features of the speech such as voice timbre and prosody characteristics are used. Voice timber is concerned with spectral properties of the vocal tract during phonation, whereas prosody relates to suprasegmental characteristics i.e., pitch, amplitude, stress, and duration. Various VC challenges (VCC) have been held to encourage the development of VC techniques and improve the quality of converted speech [198-200]. The earlier VCC aimed to convert source speech to target speech by using non-parallel and parallel data [198, 199]. Whereas, the latter [200] focused on the development of cross-lingual VC techniques, where the source speech is converted to sound like target speech using nonparallel training data and across different languages.

The earlier studies VC techniques are based on spectrum mapping using paired training data, where speech samples from both the source and target speaker uttering the same linguistic content are required for conversion. Methods using GMM [201, 202], partial least square regression [203], exemplar-based [204] techniques and others [205-207] are proposed for parallel spectral modeling. These [201-204] are "shallow" VC methods that transform source speech spectral features directly in the original feature space. Nakashika et al. [205] proposed a speaker-dependent sequence modeling method based on RNN to capture temporal correlation in an acoustic sequence. In [206, 207] deep bidirectional LSTM (DBLSTM) is employed to capture long-range contextual information and generates high-quality converted speech. DNN based methods [205-207] efficiently learn feature representation for feature mapping in parallel VC. However, require large-scale paired source and target speaker utterance data for parallel training that is not feasible for practical applications in the real world.

The VC methods for non-parallel (unpaired) training data are proposed to achieve VC for multiple speakers with different languages. The powerful VC techniques based on neural network [208], vocoder [209, 210], GAN [211-217], VAE [218-220] are introduced for non-parallel spectral modeling. Auto-encoder-based approaches attempt to learn disentangle speaker information from linguistic content and independently convert the speaker identity. Work in [220] investigates the quality of learned representation by comparing different auto-encoding methods. It was shown that a combination of Vector Quantized VAE and WaveNet [52] decoder better preserves speaker invariant linguistic content and retrieves information discarded by the encoder. However, VAE/GAN-based methods over smooth the transformed features as due to dimensionality reduction bottleneck certain low-level information, e.g. pitch contour, noise, and channel data is lost that results in the buzzy-sounding converted voices.

Recently GAN-based approaches, such as CycleGAN [211-214], VAW-GAN [215], and StarGAN [216] attempt to achieve high-quality transformed speech using non-parallel training data. Studies [212, 216] demonstrate state-of-the-art performance for multilingual VC in terms of both naturalness and similarity. However, performance is speaker-dependent and degrades for unseen speakers. Neural vocoders have rapidly become the most popular vocoding approach for speech synthesis due to their ability to generate human-like speech [193]. The vocoder learns to generate audio waveform from acoustic features. The study [210] analyzed the performance of different vocoders and showed that parallel-WaveGAN [221] can effectively simulate the data distribution of human speech with acoustic characteristics for VC. However, the performance is still restricted for unseen speaker identity and noisy samples [179]. The recent VC methods based on TTS like AttS2S-VC [222], Cotatron [223], and VTN [224] use text labels to synthesize speech directly by extracting aligned linguistic characteristics from the input voice. This assures that the converted speaker and the target speaker identity are the same. However, these methods necessitate the use of text labels, which are not always readily accessible.

Recently, one-shot VC techniques [225, 226] have been presented. In contrast to earlier techniques, the data samples of source and target speakers are not required to be seen during training. Furthermore, just one utterance from the source and target speakers is required for conversion. The speaker embedding is extracted from the target speech which can control the speaker identity of the converted speech independently. Despite these advancements, the performance of few-shot VC techniques for unseen speakers is not stable [227]. This is primarily due to the inadequacy of speaker embedding extracted from a single speech of an unseen speaker [228] that significantly impacts the reliability of one-shot conversions. The other work [229-231] adopt zero-shot VC, the source and target speakers are unseen during training also without re-training the model by employing encoder-decoder architecture. The speaker encoder extracts style and content information into style embedding and content embedding, the decoder constructs speech sample by combining style and content embedding. The zero-shot VC scenario is attractive because no adaption data or parameters are required. However, the adaptability quality is insufficient, especially when the target and source speakers are unseen, very diverse, and noisy [227]. The summary of voice conversion techniques discussed above is presented in Table 14.

Table 14: An overview of the state-of-the-art voice conversion techniques

| Methods | Technique | Features | Dataset | Limitations |
|---|---|---|---|---|
| Ming et al. [206] | DBLSTM | F0 and energy contour | ▪ CMU-ARCTIC [232] | ▪ Require parallel training data |
| Nakashika et al. [205] | Recurrent temporal restricted Boltzmann machines (RTRBMs) | MCC, F0, and aperiodicity | ▪ ATR Japanese speech database [233] | ▪ Lack temporal dependencies of speech sequences |
| Sun et al. [207] | DBLSTM-RNN | MCC, F0 and Aperiodicity | ▪ CMU-ARCTIC [232] | ▪ Require parallel training data |
| Wu et al. [14] | DBLSTM- i-vectors | 19D-MCCs, Delta and Delta-Delta, F0, 400-D i-vector | ▪ VCTK corpus | ▪ Computationally expensive |
| Liu et al. [209] | WaveNet vocoder | MCC and F0 | ▪ VCC 2018 | ▪ Performance degrades on inter-gender conversions |
| Kaneko et al. [213] | Encoder-decoder with GAN | 34D-MCC, F0, and aperiodicity | ▪ VCC 2018 | ▪ Computationally Expensive<br>▪ Domain-specific voice |
| Kameoka et al. [216] | Encoder-decoder with GAN | 36D-MCC, F0, and aperiodicity | ▪ VCC 2018 | ▪ Performance degrades on cross-gender conversion<br>▪ Limited performance for unseen speaker |

| Zhang et al. [217] | VAW-GAN | STRAIGHT spectra [234], F0 and aperiodicity | - VCC2016 | - Lack target speaker similarity |
|---|---|---|---|---|
| Huang et al. [218] | Encoder-decoder | STRAIGHT spectra [234] MCCs | - VCC 2018 | - Lack multi-target VC<br>- Introduce abnormal fluctuations in generated speech |
| Chorowski et al. [220] | VQ-VAE, WaveNet decoder | 13D-MFCC | - LibriSpeech<br>- ZeroSpeech 2017 | - Over smooth and low naturalness in generated speech<br>- Increased training complexity |
| Tanaka et al. [222] | BiLSTM encoder-LSTM decoder | Acoustic features | - CMU Arctic database | - Requires extensive training data |
| Park et al. et al. [223] | Encoder-decoder | Mel-spectrogram | - LibriTTS<br>- VCTK dataset, | - Requires transcribed data<br>- Lack target speaker similarity |
| Huang et al. [224] | VAE-vocoder | MCCs, log F0, and aperiodicity | - CMU ARCTIC dataset<br>- VCTK corpus | - Requires parallel training data |
| Lu et al. [225] | Aattention mechanism in encoder-decoder | 13D-MFCCs, PPGs and log F0 | - VCTK corpus | - Low target similarity and naturalness in generated speech |
| Liu et al. [226] | Encoder and DBLSTM | 19 MFCCs, log F0 and PPG | - VCTK corpus | - Low target similarity and naturalness in generated speech |
| Chou et al. [229] | Attention mechanism in encoder-decoder | 19 MFCCs, log F0 and PPG | - VCTK Corpus | - Low quality of converted voices in case of noisy samples |
| Qian et al. [230] | Encoder-decoder | speech spectrogram | - VCTK corpus | - Prosody flipping between the source and the target.<br>- Not well-generalized to unseen data |

**Audio Deepfake Detection**
Due to recent advances in TTS [52, 193] and VC [227] techniques, audio deepfakes have become an increased threat to voice biometric interfaces and society [13]. In the field of audio forensics, there are several approaches for identifying various types of audio spoofing. However, existing works fail to fully tackle the detection of synthetic speech [235]. In this section, we have reviewed the approaches proposed for the detection of audio deepfakes.

**Techniques based on handcrafted Features:** Yi et al. [236] presented an approach to identify TTS-based manipulated audio content. In [236] hand-crafted features Constant Q cepstral coefficients (CQCC) were to train GMM and LCNN classifier to detect TTS synthesized speech. The approach exhibits better detection performance for fully synthesized audio however performance degrades rapidly for partial synthesized audio clips. Li et al. [237] proposed a modified ResNet model Res2Net. They evaluated the model using different acoustic features and obtained the best performance with CQT features. This model exhibits better audio manipulation detection performance however generalization ability needs further improvement. In [238] Mel-spectrogram features with ResNet-34 were employed to detect spoofed speech. This approach works well however performance needs improvement. Monteiro et al. [239] proposed an ensemble-based model for the detection of synthetic speech. Deep learning models LCNNs and ResNets were used to compute the deep features which were later fused to differentiate between real and spoofed speech. This model is robust to fake speech detection, however, needs evaluation on some standard dataset. Gao et al. [240] proposed a synthetic speech detection approach based on inconsistencies. They employed a global 2D-DCT feature to train a residual network to detect the manipulated speech. The model has better generalization ability, however, the performance degrades on noisy samples. Zhang et al. [241] proposed a model to detect fake speech by using a ResNet model with a transformer encoder (TEResNet). Initially, a transformer encoder was employed to compute contextual representations of the acoustic keypoints by considering correlation between audio signal frames. The computed keypoints were then used to train a residual network to differentiate real and manipulated speech. This work shows better fake audio detection performance, however, requires extensive training data. Das et al. [242] proposed a method to detect manipulated speeches. Initially, a signal companding technique for data augmentation was used to increase the diversity of training data. Then CQT features were computed from the obtained data which were later used to train the LCNN classifier. The method improves the fake audio detection accuracy however requires extensive training data.

Aljasem et al. [12] proposed a hand-crafted features-based approach to detect cloned speeches. Initially, sign-modified acoustic local ternary pattern features were extracted from input samples. Then the computed keypoints were used to train an asymmetric bagging-based classifier to categorize the bonafide and fake speeches. The work is robust to noisy cloned voice replay attacks, however, performance needs further improvement. Ma et al. [243] presented a continual learning-based technique to enhance the generalization ability of manipulated speech detection system. A knowledge distillation loss function was introduced in the framework to enhance the learning ability of the model. The approach is computationally efficient and can detect unseen fake spoofing manipulations, however performance is not evaluated on noisy samples. Borrelli et al. [244] employed bicoherence features together with long-term short-term features. The extracted features were used to train three different types of classifiers i.e., random forest, a linear SVM, and radial basis function (RBF) SVM. The method obtains the best accuracy with the SVM classifier. However, due to handcrafted features, this work is not generalized to unseen manipulations. In [245] bispectral analysis was performed to identify specific and unusual spectral correlations present in GAN generated speech samples. Similarly in [246] bispectral and Mel-cepstral analysis was performed to detect missing durable power components in synthesized speech. The computed features were used to train several ML-based classifiers and attained the best performance using Quadratic SVM. These approaches [245, 246] are robust to TTS synthesized audio, however may not detect high-quality synthesized speech. Malik et al. [247] proposed a CNN for cloned speech detection. Initially, audio samples were converted to spectrograms on which a CNN framework was used to compute deep features and classify real and fake speech samples. This approach shows better fake audio detection accuracy however, performance degrades on noisy samples. Chen et al. [248] proposed a DL-based framework for audio deepfakes detection. The 60-dimensional linear filter banks (LFB) were extracted from speech samples that were later used to train a modified ResNet model. This work improves the fake audio detection performance, however, suffers from high computational cost. Huang et al. [249] presented an approach for audio spoofing detection. Initially, short-term zero-crossing rate and energy were utilized to identify the silent segments from each speech signal. In the next step, the linear filter bank (LFBank) key-points were computed from the nominated segments in the relatively high-frequency domain. Lastly, an attention-enhanced DenseNet-BiLSTM framework was built to locate audio manipulations. This method [249] can avoid over-fitting, however, it is at the expense of high computational cost. Wu et al. [250] introduced a novel key-points genuinization based light convolutional neural networks (LCNN) framework for the identification of synthetic speech manipulation. The attributes of the original speech were utilized to train a model using CNN. It was then converted to an original key-point distribution closer to that of genuine speech. The transformed key-points were used with an LCNN to identify genuine and altered speech. This approach [250] is robust to synthetic speech manipulation detection. It is, however, unable to deal with replay attack detection.

**Techniques based on Deep Features:** Zhang et al. [251] proposed a DL-based approach using ResNet-18 and one-class (OC) softmax. They trained the model to learn feature space in which real speech can be discriminated from manipulated samples by a certain margin. This method improves the performance generalization ability against unseen attacks, however, performance degrades on VC attacks generated using waveform filtering. In [252] authors proposed a Light Convolutional Gated RNN (LCGRNN) model to compute the deep features and classify the real and fake speech. This model is computationally efficient however, not generalized well to real-world examples. Hua et al. [253] proposed end-to-end synthetic speech detection model Res-TSSDNet for the computation of deep features and classification. This model is generalized well to unseen samples however at the expense of increased computational cost. Wang et al. [254] proposed a DNN based approach with a layer-wise neuron activation mechanism to differentiate between real and synthetic speech. This approach performs well for fake audio detection, however, the framework requires evaluation on challenging datasets. Jiang et al. [255] proposed a self-supervised learning-based approach comprising eight convolutional layers to compute the deep features and classify the original and fake speeches. This work is computationally efficient, however detection accuracy needs enhancement.

Most of the above mentioned fake speech detection have been evaluated on ASVspoof2019[180], however, the recently launched ASVspoof2021[256] has opened new challenge for the research community. This dataset has introduced a separate speech deepfake category that includes highly compressed TTS and VC samples without speaker verification.

Table 15: An overview of Audio deepfake detection techniques

| Author | Technique | Features | Best Evaluation performance | Dataset | Limitations |
|---|---|---|---|---|---|
| Hand-crafted features | | | | | |
| Li et al. [237] | Res2Net | CQT | EER=2.502 | ASVspoof2019 | ▪ Needs generalization improvement |
| Yi et al. [236] | GMM/LCNN | CQCC | EER=19.22 (GMM) EER=6.99 (LCNN) | Propriety | ▪ Performance degrades for partial synthesized audio clip |

| Das et al. [242] | LCNN | CQT | EER=3.13 | ASVspoof2019 | ▪ Requires extensive training data |
| Aljasem et al. [12] | Asymmetric bagging | Combination of MFCC, GTCC, ALTP, and spectral features | EER=5.22 | ASVspoof2019 | ▪ Performance needs further improvement |
| Ma et al. [243] | CNN | 60-D LFCC | EER=9.25 | ASVspoof2019 | ▪ Performance degrades on noisy samples |
| AlBadawy et al. [245] | logistic regression classifier | Bispectral features | AUC=0.99 | Propriety | ▪ Performance may degrade on high quality speech samples |
| Singh et al. [246] | Quadratic SVM | Bispectral and mel-cepstral features | Acc=96.1% | Propriety | ▪ Needs evaluation on a large scale dataset |
| Gao et al. [240] | ResNet | 2D-DCT features | EER=4.03 | ASVspoof2019 | ▪ Performance degrades on noisy samples |
| Aravind et al. [238] | ResNet34 | Mel-spectrogram features | EER=5.87 | ASVspoof2019 | ▪ Performance needs improvement |
| Monteiro et al. [239] | LCNN/ResNet | Spectral features | EER=6.38 | Propriety | ▪ Results should be evaluated on some standard dataset |
| Chen et al. [248] | ResNet | 60-dimensional LFB | EER=1.81 | ASVspoof2019 | ▪ Computationally expensive approach |
| Huang et al. [249] | DenseNet-BiLSTM | LFBank | EER=0.53 | ASVspoof 2019 | ▪ Computationally complex approach. |
| Wu et al. [250] | LCNN | Genuine speech features | EER= 4.07 | ASVspoof 2019 | ▪ Can't deal with replay attack detection. |
| Zhang et al. [241] | TEResNet | Spectrum features | EER=5.89 | ASVspoof2019 | ▪ Requires extensive training data |
| | | | EER=3.99 | Fake-or-Real dataset [257] | |
| **Deep Learning features** | | | | | |
| Zhang et al. [251] | ResNet-18+OC-softmax | Deep features | EER=2.19 | ASVspoof2019 | ▪ Performance degrades on VC. |
| Gomez-Alanis et al. [252] | LCG- RNN | Deep features | EER=6.28 | ASVspoof 2019 | ▪ Fails to generalize for unseen attacks |
| Hua et al. [253] | Res-TSSDNet | Deep features | EER=1.64 | ASVspoof2019 | ▪ Computationally complex |
| Jiang et al. [255] | CNN | Deep features | EER=5.31 | ASVspoof2019 | ▪ Performance needs further improvement |
| Wang et al. [254] | DNN | Deep features | EER=0.021 | Fake-or-Real dataset [257] | ▪ Requires evaluation on challenging dataset |

## 4.7 Discussion

This section provides a summary of recent significant advancements in audio-visual deepfake creation and detection techniques.

**Generation**

In recent years, the deepFake generation has advanced significantly. The high quality of generated images across different visual manipulation categories (face-swap, face-reenactment, lip-sync, entire face synthesis, and attribute manipulation) has made it increasingly difficult for human eyes to differentiate between fake and genuine content. Among the significant advances are (1) unpaired self-supervised training strategies to avoid the requirement for extensive labeled training data, and (2) addition of AdaIN layers, pix2pixHD network, self-attention modules, and feature disentanglement for improved synthesized faces (3) one/few-shot learning strategies to enable identity theft with limited target training data (4) use of temporal discriminator and optical flow estimation to improve coherence in the synthesized videos (5) introduction of secondary network for seamless blending of composites to reduce the boundary artifacts (6) use of multiple loss functions to handle different tasks such as conversion, blending, occlusion, pose, illumination, etc., for improved final output and (7) adoption of perceptual loss with pre-trained VGG-Face network dramatically enhanced synthesize facial quality. Current deepfake systems have a few limitations such as in facial reenactment generation techniques, frontal poses are always used to drive and create the content. As a result, the reenactment is restricted to a somewhat static performance. Currently, Face-swapping onto the body of lookalike is performed to achieve facial reenactment, however, this approach has limited flexibility because having a good match is not always achievable. Moreover, face reenactment depends on the driver's performance to portray the target identity personality. Recently, there has been a trend towards identity-independent deepfakes generation models. Another development is real-time deepfakes that allows swapping faces in video chats. Real-time deepfakes at 30fps have been achieved in works such as [64, 102]. The next generation deepfakes are expected to utilize video stylization techniques to generate target manipulated content with projected expression and mannerism. Although, existing deepfakes are not perfect, however, the rapid development of high-quality real/fake image dataset promote the deepfake generation research.

In recent years, the quality of synthetic voice has significantly improved by using deep learning techniques. The significant improvements include voice adaptation [55] [191], one/few-shot learning [225, 226], self-attention network [229], and cross-lingual voice transfer [212, 216]. However, their ability to produce more human-like natural-sounding utterances with limited training samples under varying settings remains challenging [258].

**Detection**

In this subsection, we have presented a summary of the work performed for audiovisual deepfakes detection. Based on the in-depth analysis of various detection approaches, we have concluded that most of the existing detection work is based on employing DL-based approaches and showing robust performance close to 100%. The main reason for the accurate detection accuracy of models is due to the presence of fingerprint information, visible artifacts in the audiovisual manipulated samples. However, now, the researchers have presented such approaches which have removed the information from the forged samples while maintaining the fake realism which is showing a new challenge even for the high-performing attribute manipulation detection frameworks. It has been observed that most of the existing detection techniques perform well on face swap detection, and are relatively easier to identify as entire face is swapped with target identity which usually leaves artifacts. However, expression swap and lip-sync are more challenging to detect as these manipulations tamper soft biometrics of the same person identity. In the case of visual manipulation detection, most of the research work has utilized ACC and AUC for the evaluation of their results, while audio deepfakes detection has used the EER metric. For visual deepfakes detection, it has been observed that it's easy for the research community to detect image-based manipulations in comparison to video-based deepfakes. While for audio manipulations VC detection is more challenging as compared to TTS. Both for audio or visual deepfakes, most of the research work have use publically available datasets instead of using their own synthesized datasets. The existing work has reported robust performance for audiovisual deepfakes detection, however, has faced serious performance drop for unseen cases which depicts a lack of generalization ability. Moreover, these approaches are unable to give proof to differentiate real and manipulated content, so, these approaches lack explainability. It has been observed that several deepfake detection methods are presented in previous years, however, due to implementation complexities such as variation in datasets, configuration environment, and complicated architectures, it is difficult to implement and use them. Now, different software and online platforms such as DeepFake-o-meter [259], FakeBuster [260], and Video Authenticator (not publicly available) [261] are introduced to ease the audio-visual detection and give access to the general audience. However, these platforms are at the infancy stage and need further development to handle emerging deepfakes.

We have used a figure representation to group the existing work performed for audio and visual deepfake detection (Fig. 10). Table 16 presents the detailed description of each category. Existing approaches have either targeted spatial and temporal artifacts left during the generation, or data-driven classification. The spatial artifacts include inconsistencies [75, 78, 110, 245, 262-264], abnormalities in background [155, 265, 266], and GAN fingerprints [71, 157, 267]. The temporal artifacts involve detecting variation in a person's behavior [79, 87, 268], physiological signals [74, 75, 83, 89], coherence [269, 270], and video frame synchronization [72, 82, 94, 134]. Instead of focusing on a specific artifact, some approaches are data-driven, which detect manipulations by classification [70, 80, 84, 86, 95-98, 114, 118, 138, 156, 158, 250, 254, 271-274] or anomaly identification [116, 117, 275-277]. Moreover, in Fig, red-colored references are showing the DL-based approaches employed for deepfakes detection, while others show the hand-coded methods.

**Table 16: Description of classification categories for existing deepfake detection methods**

| | |
|---|---|
| **Inconsistencies** | Visible artifacts within the frame such as inconsistent head poses and landmarks etc. |
| **Environment** | Abnormalities in the background such as lighting and other details. |
| **Forensics** | GAN fingerprints left during the generation process. |
| **Behavioral** | Monitoring abnormal gestures and facial expressions. |
| **Synchronization** | Temporal consistency such as inconsistencies between adjacent frames/modality. |
| **Physiology** | Lack of biological signals such as eye blinking patterns and heart rate |
| **Coherence** | Missing optical flow field and artifacts such as flickering and jitter between frames |
| **Classification** | End-to-end CNN based data-driven models |
| **Anomaly Detection** | Outliers identification such as reconstructing real images and comparing to the encoded image. They are used to see unknown creation methods. |

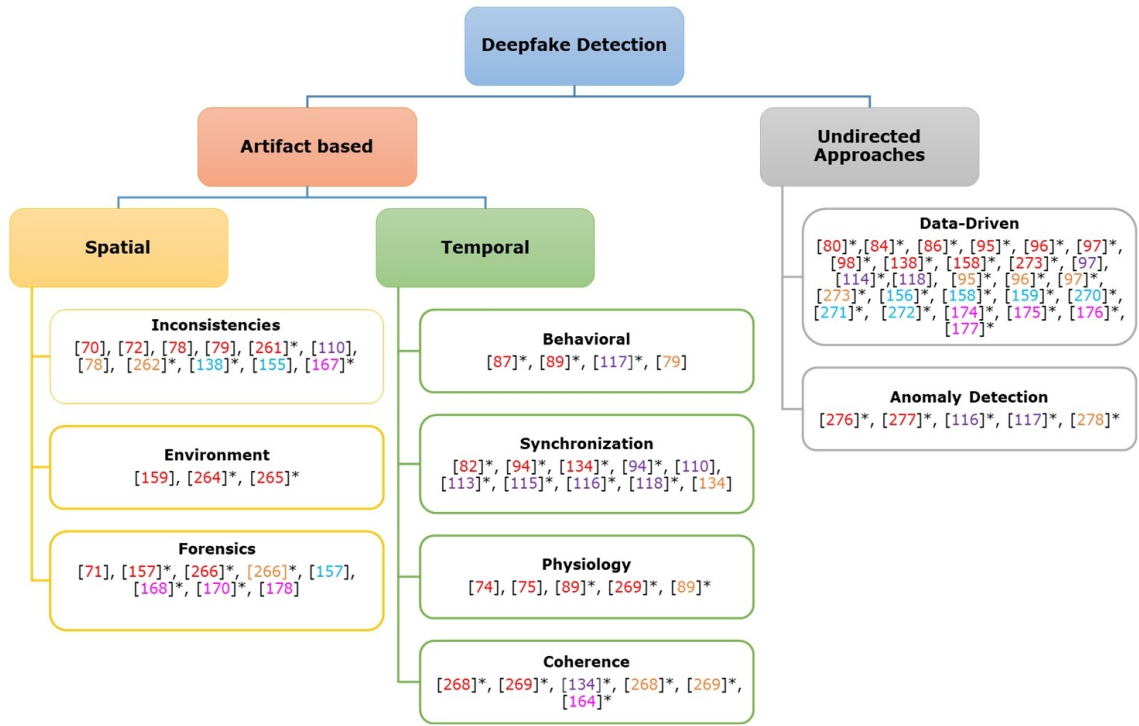

**Figure 10:** Categorization of deepfake detection techniques (The red color shows Face-Swap detection approaches, purple for Face-Reenactment, Orange for lip-syncing, Blue for facial image synthesis, and pink for facial attribute manipulation detection techniques, where * shows deep-learning based approaches)

## 4.8 Open Challenges in Deepfakes Generation

Although extensive efforts have been shown to improve the visual quality of generated deepfakes there are still several challenges that need to be addressed. A few of them are discussed below.

**Generalization:** The generative models are data-driven, and therefore they reflect the learned features during training in the output. To generate high-quality deepfakes a large amount of data is required for training. Moreover, the training process itself requires hours to produce convincing deepfake audiovisual content. Usually, it is easier to obtain a dataset of the driving content but the availability of sufficient data for a specific victim is a challenging task. Also retraining the model for each specific target identity is computationally complex. Because of this, a generalized model is required to enable the execution of a trained model for multiple target identities unseen during training or with few training samples available.

**Identity Leakage**: The preservation of target identity is a problem when there is a significant mismatch between the target identity and the driving identity, specifically in face reenactment tasks where target expressions are driven by some source identity. The facial data of the driving identity is partially transferred to the generated face. This occurs when training is performed on single or multiple identities, but data pairing is accomplished for the same identity.

**Paired Training:** A trained supervised model can generate high-quality output but at the expense of data pairing. Data pairing is concerned with generating the desired output by identifying similar input examples from the training data. This process is laborious and inapplicable to those scenarios where different facial behaviors and multiple identities are involved in the training stage.

**Pose Variations and Distance from camera**: Existing deepfake techniques generate good results of the target for frontal facial view. However, the quality of manipulated content degrades significantly for scenarios where a person is looking off camera. This results in undesired visual artifacts around the facial region. Furthermore, another big challenge for convincing deepfake generation is the facial distance of the target from the camera, as an increase in distance from capturing devices results in low-quality face synthesis.

**Illumination Conditions**: Current deepfake generation approaches produce fake information in a controlled environment with consistent lighting conditions. However, an abrupt change in illumination conditions such as in indoor/outdoor scenes results in color inconsistencies and strange artifacts in the resultant videos.

**Occlusions**: One of the main challenges in deepfake generation is the occurrence of occlusion, which results when the face region of the source and victim are obscured with a hand, hair, glasses, or any other items. Moreover, occlusion can be the result of the hidden face or eye portion which eventually causes inconsistent facial features in the manipulated content.

**Temporal Coherence**: Another drawback of generated deepfakes is the presence of evident artifacts like flickering and jitter among frames. These effects occur because the deepfake generation frameworks work on each frame without taking into account the temporal consistency. To overcome this limitation, some works either provide this context to generator or discriminator, consider temporal coherence losses, employ RNNs, or take a combination of all these approaches.

**Lack of realism in synthetic audio**: Though the quality is certainly getting much better, there is still a need for improvement. The main challenges of audio-based deepfakes are the lack of natural emotions, pauses, breathiness, and the pace at which the target speaks.

Based on the above-mentioned limitations we can argue that there exists a need to develop effective deepfake generation methods that are robust to variations in illumination conditions, temporal coherence, occlusions, pose variations, camera distance, identity leakage, and paired training.

## 4.9 Challenges in Deepfakes detection methods

Although remarkable advancements have been made in the performance of deepfake detectors there are numerous concerns about current detection techniques that need attention. Some of the challenges of deepfake detection approaches are discussed in this section.

**Quality of Deepfake Datasets:** The accessibility of large databases of deepfakes is an important factor in the generation of deepfake detection techniques. However, analyzing the quality of videos from these datasets reveals several ambiguities in comparison to actual manipulated content found on the internet. Different visual artifacts that can be visualized in these databases are: i) temporal flickering in some cases during the speech, ii) blurriness around the facial regions, iii) over smoothness in facial texture/lack of facial texture details, iv) lack of head pose movement or rotation, v) lack of face occluding objects such as glasses, lightning effect, etc., vi) sensitive to variations in input posture or gaze, skin color inconsistency, and identity leakage, and vii) limited availability of a combined high-quality audio-visual deepfake dataset. The aforementioned dataset ambiguities are due to imperfect steps in the manipulation techniques. Furthermore, manipulated content of low quality can be barely convincing or create a real impression. Therefore, even if detection approaches exhibit better performance over such videos it is not guaranteed that these methods will perform well when employed in the wild.

**Performance Evaluation:** Presently, deepfake detection methods are formulated as a binary classification problem, where each sample can be either real or fake. Such classification is easier to build in a controlled environment, where we generate and verify deepfake detection techniques by utilizing audio-visual content that is either original or fabricated. However, for real-world scenarios, videos can be altered in ways other than deepfakes, so content not detected as manipulated does not guarantee the video is an original one. Furthermore, deepfake content can be the subject of multiple types of alteration i.e. audio/visual, and therefore a single label may not be completely accurate. Moreover, in visual content with multiple people's faces, usually, one or more of them are manipulated with deepfakes over a segment of frames. Therefore, the binary classification scheme should be enhanced to multiclass/multi-label and local classification/detection at the frame level, to cope with the challenges of real-world scenarios.

**Lack of Explainability in Detection Methods:** Existing deepfake detection approaches are typically designed to perform batch analysis over a large dataset. However, when these techniques are employed in the field by journalists or law enforcement, there may only be a small set of videos available for analysis. A numerical score parallel to the probability of an audio or video being real or fake is not as valuable to the practitioners if it cannot be confirmed with appropriate proof of the score. In those situations, it is very common to demand an explanation for the numerical score for the analysis to be believed before publication or utilization in a court of law. Most deepfakes detection methods lack such an explanation, however, particularly those which are based on DL approaches due to their black-box nature.

**Lack of fairness and Trust:** It has been observed that existing audio and visual deepfakes datasets are biased and contain imbalanced data of different races and genders. Furthermore, the employed detection techniques can be biased as well. Although researchers have started doing work in this area to fill this gap, however, very little work is available [278]. Hence, there is an urgent need to introduce such approaches that improve the data and fairness in detection algorithms.

**Temporal Aggregation:** Existing deepfake detection methods are based on binary classification at the frame level, i.e. checking the probability of each video frame as real or manipulated. However, these approaches do not consider temporal consistency between frames, and suffer from two potential problems: (i) deepfake content shows temporal artifacts, and (ii) real or fake frames could appear in sequential intervals. Furthermore, these techniques require an extra step to compute the integrity score at the video level, as these methods need to combine the score from each frame to generate a final value.

**Social Media Laundering:** Social platforms like Twitter, Facebook, or Instagram are the main online networks used to spread audio-visual content among the population. To save the bandwidth of the network or to secure the user's privacy, such content is stripped of meta-data, down-sampled, and substantially compressed before uploading. These manipulations, normally known as social media laundering, remove clues with respect to underlying forgeries and eventually increase false positive detection rates. Most deepfake detection approaches employing signal level key-points are more affected by social media laundering. A measure to increase the accuracy of deepfake identification approaches over social media laundering is to keenly include simulations of these effects in training data, and also increase the evaluation databases to contain data on social media laundered visual content.

**DeepFake Detection Evasion:** Mostly deepfake detection methods are concerned to locate missing information and artifacts left during the generation process. However, detection techniques may fail in case of the unavailability of such data as attackers attempt to remove such traces during the manipulation generation process. Such fooling techniques are classified into three types such as adversarial perturbation attack, elimination of manipulation traces in the frequency domain, and employing image filtering to mislead detectors. In the case of visual adversarial attacks, different perturbations such as random cropping, noise, and JPEG compression, etc., are added to the training data, which ultimately results in high false alarms for detection methods. Different works [279, 280] have evaluated the performance of state-of-the-art visual deepfake detectors under the presence of adversarial attack and showed the intense reduction in accuracy. While in the case of audio, studies such as [281, 282] showed that several adversarial pre/post-processing operations can be used to evade spoof detection. Similarly, the second method is concerned with improving the quality of GAN-generated samples by enhancing spectral distributions [283]. Such methods ultimately result in removing fake traces in the frequency domain and complicates the detection process [284, 285]. The third method uses advanced image filtering techniques to improve generation quality such as removal of fingerprints left during generation and addition of noise to remove fake signs [286-288]. The aforementioned methods impose a real challenge for deepfake detection methods, thus research community needs to propose such techniques that are robust and reliable to such attacks.

## 5  Deepfake Datasets

To analyze the detection accuracy of proposed methods it is of utmost importance to have a good and representative dataset for performance evaluation. Moreover, the techniques should be validated over cross datasets to show their generalization power. Therefore, researchers have put in significant effort over the years by preparing the standard datasets for manipulated visual and audio content. In this section, we have presented a detailed review of the standard datasets that are currently used to evaluate the performance of audio and video deepfake detection techniques. Tables 17 and 18 show a comparison of available video and audio deepfake datasets respectively.

### 5.1 Video Datasets

**UADFV**: The first dataset released for deepfake detection was UADFV [71]. It consists of a total of 98 videos, where 49 are real videos collected from YouTube and manipulated by using the FakeApp application [41] to generate 49 fake videos. The average length of videos is 11.14 sec with an average resolution of 294×500 pixels. However, the visual quality of videos is very low, and the resultant alteration is obvious and thus easy to detect.

**DeepfakeTIMIT:** DeepfakeTIMIT [271] is another standard dataset for deepfake detection which was introduced in 2018. This dataset consists of a total of 620 videos of 32 subjects. For each subject, there are 20 deepfake videos of two quality levels, where 10 videos belong to DeepFake-TIMIT-LQ and the remaining 10 belong to DeepFake-TIMIT-HQ. In DeepFake-TIMIT-LQ, the resolution of the output image is 64×64, whereas, in DeepFake-TIMIT-HQ, the resolution of output size is 128×128. The fake content is generated by employing face swap-GAN [62], however, the generated videos are only 4 seconds long and the dataset contains no audio channel manipulation. Moreover, the resultant videos are often blurry and people in actual videos are mostly presented in full frontal face view with a monochrome color background.

**FaceForensics++**: One of the most famous datasets for deepfake detection is FF++ [98]. This dataset was presented in 2019 as an extended form of the FaceForensics dataset [289], which contains videos with facial expressions manipulation only, and which was released in 2018. The FF++ dataset has four subsets named FaceSwap [290], DeepFake [42], Face2Face [36], and NeuralTextures [291]. It contains 1000 original videos collected from the YouTube-8M dataset [292] and 3,000 manipulated videos generated using the computer graphics and deepfake approaches specified in [289]. This dataset is also available in two quality levels i.e. uncompressed and H264 compressed format, which can be used to evaluate the performance of deepfake detection approaches on both compressed and uncompressed videos. The FF++ dataset fails to generalize lip-sync deepfakes however, and some videos exhibit color inconsistencies around the manipulated faces.

**Celeb-DF**: Another popular dataset used for evaluating deepfake detection techniques is Celeb-DF [265]. This dataset presents videos of higher quality and tries to overcome the problem of visible source artifacts found in previous databases. The CelebDF dataset contains 408 original videos and 795 fake videos. The original content was collected from Youtube, which is divided into two parts named Real1 and Real2 respectively. In Real1, there are a total of 158 videos of 13 subjects with different gender and skin color. Real2 comprises 250 videos, each having a different subject, and the synthesized videos are generated from these original videos through the refinement of existing deepfake algorithms [293, 294].

**Deepfake Detection Challenge (DFDC)**: Recently, the Facebook community launched a challenge, aptly named the Deepfake Detection Challenge (DFDC)-preview [295], and released a new dataset that contains 1131 original videos and 4119 manipulated videos. The altered content is generated using two unknown techniques. The final version of the DFDC database is publicly available on [296]. It contains 100,000 fake videos along with 19,000 original samples. The dataset is created using various face-swap-based methods with different augmentations (i.e., geometric and color transformations, varying frame rate, etc.) and distractors (overlaying different types of objects) in a video.

**DeeperForensics (DF):** Another Large-Scale dataset for deepfake detection containing 50,000 original and 10,000 manipulated videos is built-in [297]. A novel conditional autoencoder, namely DF-VAE is used to create manipulated videos. The dataset comprises highly diverse samples in terms of actor's appearance. Further, a mixture of distortions and perturbations such as compression, blurry, noise, etc. are added to better represent the real-world scenarios. As compared to previous datasets [71, 265, 271], the quality of generated samples is significantly improved.

**WildDeepfake:** WildDeepfake (WDF)[298] is considered as one of the challenging deepfake detection datasets. It contains both real and deepfake samples collected from the internet in comparison to existing datasets.

All of the above-mentioned datasets contain a synthesized face portion only and the datasets lack upper/full body deepfakes. A more robust dataset is needed which should be able to synthesize the entire body of the source person.

Table 17: Comparison of Deepfakes detection datasets

|  | UADFV [71] | DF-TIMIT[271] | FF++ [98] | Celeb-DF [265] | DFDC-preview [296] | DF [297] | WDF[298] |
|---|---|---|---|---|---|---|---|
| **Released** | Nov, 2018 | Dec, 2018 | Jan, 2019 | Nov, 2019 | Oct, 2019 | June, 2020 | Oct, 2020 |
| **Total videos** | 98 | 620 | 4000 | 1203 | 5250 | 60,000 |  |
| **Real content** | 48 | Nill | 1000 | 408 | 1131 | 10,000 | 3,805 |
| **Fake content** | 48 | 620 | 3000 | 795 | 4119 | 50,000 | 3,509 |
| **Tool/ technology used for fake content generation** | FakeApp application [41] | faceswap-GAN [62] | deepfake, CG-manipulations | deepfake | Unknown | DF-VAE | unknown |
| **Avg. Duration** | 11.4 sec | 4 sec | 18 sec- | 13 sec | 30 sec | - | - |
| **Resolution** | 294×500 | 64×64 (LQ) 128×128 (HQ) | 480p, 720p, 1080p | various | 180p – 2160p | 1920×1080 | various |
| **Format** | - | JPG | H.264, CRF=0, 23, 40 | mp4 | H.264 | mp4 | mp4 |
| **Visual quality** | low | low | low | high | high | high | high |
| **Temporal flickering** | yes | yes | yes | improved | improved | Significantly improved | - |
| **modality** | visual | Audio/visual | visual | visual | Audio/visual | visual | visual |

## 5.2 Audio Datasets

**LJ speech and M-AILabs dataset:** LJSpeech [299] and M-AILabs [300] dataset are famous for the real-speech database employed in numerous TTS applications, i.e. DeepVoice 3 [193]. The LJSpeech database is comprised of 13,100 clips totaling 24 hours length. All utterances are recorded by a female speaker. The M-AILABS dataset consists of total 999 hours and 32 minutes of audio. This dataset was created with multiple speakers in 9 different languages.

**Mozilla TTS:** Mozilla Firefox a well-known publicly available browser, released the biggest open-source database of people speaking [301]. Initially, the database included 1400 hours of recorded voices, in 18 different languages, in 2019. Later it was extended to 7,226 hours of recoded voices in 54 diverse languages. This dataset contains 5.5 million audio clips and was employed by Mozilla's Deep Speech toolkit.

**ASV spoof 2019:** Another well-known dataset for fake audio detection is ASVspoof-2019 [180], which is comprised of two parts for performing logical access (LA) and physical access (PA) state analysis. Both LA and PA are created from the VCTK base corpus, which comprises audio clips taken from 107 speakers (46 males, 61 females). LA consists of both voice cloning and voice conversion samples, whereas PA consists of replay samples along with bona fide ones. Both datasets are further divided into three databases, named training, development, and evaluation, which contain clips from 20- (8 males, 12 females), 10- (4 males, 6 females), and 48- (21 males, 27 females) speakers respectively. Further categorization is diverse in terms of presenters, and the recording situations are the same for all source samples. The training and development sets contain spoofing occurrences created with the same method/conditions (labeled as known attacks), while the evaluation set contains samples with unknown attacks.

**Fake-or-Real (FOR) dataset:** The FOR database [257] is another dataset that is widely employed for synthetic voice detection. This database consists of over 195,000 samples both from humans and AI-synthetic speech. This database groups samples from the new TTS method (i.e. Deep Voice 3[193] and Google-Wavenet [52]) together with diverse human speech samples ( i.e Arctic Dataset, LJSpeech Dataset, VoxForge Dataset). The FOR database has four versions, namely for-original (FO), for-norm (FN), for-2sec (F2S), and for-rerec (FR). FO contains unbalanced voices without alterations, while FN comprises balanced unaltered samples in terms of gender, class, and volume, etc. F2S contains data from FN, however, the samples are trimmed to 2 seconds, and the FR version is a rerecorded version of the F2S database, to simulate a condition in which an invader passes a sample via a voice channel (i.e. a cellphone call or a voice message).

**Baidu Dataset:** The Baidu Silicon Valley AI Lab cloned audio dataset is another database employed for cloned speech detection [55]. This database is comprised of 10 ground truth speech recordings, 120 cloned samples, and 4 morphed samples.

Table 18: Comparison of audio fakes detection datasets

|  | LJ speech dataset [299] | M-AILabs dataset [300] | Mozilla TTS [301] | FOR dataset [257] | Baidu Dataset [55] | ASV spoof 2019[180] |
|---|---|---|---|---|---|---|
| **Released** | 2017 | 2019 | 2019 | 2019 | 2018 | 2019 |
| **Total samples** | 13100 | - | 5.5 million | 195,000 | 120 | 122157 |
| **Length (hrs)** | 24 | 999 hrs 32min | 7226 | - | 0.6 | - |
| **Speaker Accent** | Native | Native | 24% US English, 8% British English | Native | US English, British English | Native |
| **Languages** | 1 | 9 | 54 | 1 | 1 | 1 |
| **Speaker gender** | 100% Female | Male, female | 47% Male 15% Female | 50% male, 50% female | 50% male, 50% female | 43% male, 57 female |
| **Format** | wav | wav | mp3 | mp3 | mp3 | mp3 |
| **Tool/ technology used for generation** | recorded | recorded | recorded | Deep Voice 3, TTs, Google-Wavenet etc. [257] | Neural voice cloning [55] | Tacotron2 [9] and WaveNet [10] |

# 6  Future Directions

Synthetic media is gaining a lot of attention because of its potential positive and negative impact on our society. The competition between deepfake generation and detection will not end in the foreseeable future, although impressive work has been presented for the generation and detection of deepfakes. There is still, however, room for improvement. In this section, we discuss the current state of deepfakes, their limitations, and future trends.

## 6.1 Creation

Visual media has more influence compared to text-based disinformation. Recently, the research community has focused more on the generation of identity agnostic models and high-quality deepfakes. A few distinguished improvements are i) a reduction in the amount of training data due to the introduction of un-paired self-supervised methods [302], ii) quick learning, which allows identity stealing using a single image [129, 131], iii) enhancements in visual details [124, 145], iv) improved temporal coherence in generated videos by employing optical flow estimation and GAN based temporal discriminators [103], v) the alleviation of visible artifacts around face boundary by adding secondary networks for seamless blending [66], and vi) improvements in synthesized face quality by adding multiple losses with different responsibilities, such as occlusion, creation, conversion, and blending [108]. Several approaches have been proposed to boost the visual quality and realism of deepfake generation, however, there are a few limitations. Most of the current synthetic media generation focuses on a frontal face pose. In facial reenactment, for good results the face is swapped with a lookalike identity. However, it is not possible to always have the best match, which ultimately results in identity leakage.

AI-based manipulations are not restricted to the creation of visual content only, leading to a generation of highly genuine audio deepfakes. The quality of audio deepfakes has significantly improved and requires less training data in to generate more realistic synthetic audio of the target speaker. The employment of synthesized speech for impersonating targets can produce highly convincing deepfakes with a marked negative adverse impact on society. The current audio-visual content is generated separately using multiple disconnected steps, which ultimately results in the generation of asynchronous content. Present deepfake generation focuses on the face region only, however the next generation of deepfakes is expected to target full body manipulations, such as a change in body pose, along with convincing expressions. Target-specific joint audio-visual synthesis with more naturalness and realism in speech is a new cutting-edge application of the technology in the context of persona appropriation [104, 303]. Another possible trend is the creation of real-time deepfakes. Some researchers have already reported attaining real-time deepfakes at 30fps [64]. Such alterations will result in the generation of more believable deepfakes.

## 6.2 Detection

To prevent deepfakes misinformation and disinformation, some authors presented approaches to identify forensic changes made within visual content by employing the concept of blockchain and smart contracts [304, 305]. In [305] the authors utilized Ethereum smart contracts to locate and track the origin and history of manipulated information and its source, even in the presence of multiple manipulation attacks. This smart contract applied the hashes of the interplanetary file system to save videos together with their metadata. This method may perform well for deepfake identification; however, it is applicable only if the metadata of videos do exist. Thus, development and adoption of such techniques could be useful for the newswires, however, the vast majority of content created by normal citizens won't be protected by such techniques.

Recent automated deepfake identification approaches typically deal with face swapping videos, and the majority of uploaded fake videos belong in this category. Major improvements in detection algorithms include i) identification of artifacts left during the generation process, such as inconsistencies in head pose [71], lack of eye blinking [77], color variations in facial texture [155] and teeth alignment, ii) detection of unseen GAN generated samples, iii) spatial-temporal features, and iv) psychological signals like heart rate [89], and an individual's behavior patterns [79]. Although extensive work has been presented for automated detection, however, these automated detection methods are expected to be short-lived and require improvements on multiple fronts. Following are many of unresolved challenges in the domain of deepfake detection.

- The existing methods are not robust to post-processing operations like compression, noisy effects, light variations, etc. Moreover, limited work has been presented that can detect both audio and visual deepfakes.
- Recently, most of the techniques have focused on face-swap detection by exploiting its limitations, like visible artifacts. However, with immense developments in technology, the near future will produce more sophisticated face-swaps, such as impersonating someone, with the target having a similar face shape, personality, and hairstyle. Aside from this, other types of deepfake, like face-reenactment and lip-synching are getting stronger day by day.
- Existing deepfake detectors have mainly relied on the signatures of existing deepfakes by using ML techniques, including unsupervised clustering and supervised classification methods, and therefore they are less likely to detect unknown deepfakes. Both anomaly-based and signature-based detection methods have their own pros and cons. For example, anomaly detection-based approaches show a high false alarm rate because they may classify a bona fide multimedia artifact whose patterns are rare in the dataset as an anomaly. On the other hand, signature-based approaches cannot discover unknown attacks [310]. Therefore, the hybrid approach of using both anomaly and signature-based detection needs to be tried out to identify known and unknown attacks. Furthermore, a collaboration with the RL method could be added to the hybrid signature and anomaly approach. More specifically, RL can give a reward (or penalty) to the system when it selects frames of deepfakes that contain (or do not contain) anomalies, or any signs of manipulation. Additionally, in the future, deep reinforcement active learning approaches [313,314] could play a pivotal role in the detection of deepfakes.
- Anti-forensic or adversarial ML techniques can be employed to reduce the classification accuracy of automated detection methods. The game theoretic approaches could be employed to mitigate the adversarial attacks on deepfake detectors. Additionally, Reinforcement Learning (RL) and particularly deep reinforcement learning (DRL) is extremely efficient in solving intricate cyber-defense problems. Thus, DRL could offer great potential for not only deepfake detection but also to counter antiforensic attacks on the detectors. Since RL can model an autonomous agent to take sequential actions optimally with limited or without prior knowledge of the environment, thus it could be used to meet a need for developing algorithms to capture traces of anti-forensic

- processing, and to design attack-aware deepfake detectors. The defense of the deepfake detector against adversarial input could be modeled as a two-player zero-sum game with which player utilities sum to zero at each time step. The defender here is represented by an actor-critic DRL algorithm [306].
- The current deepfake detectors face challenges, particularly due to incomplete, sparse, and noisy data in training phases. There is a need to explore innovative AI architectures, algorithms, and approaches that "bake in" physics, mathematics, and prior knowledge relevant to deepfakes. Embedding physics and prior knowledge using knowledge-infused learning into AI will help to overcome the challenges of sparse data and will facilitate the development of generative models that are causal and explanative.
- Most of the existing approaches have focused on one specific type of feature, such as landmark features. However, as the complexity of deepfakes is increasing, it is important to fuse landmark, photoplethysmography (PPG) and audio-based features. Likewise, it is important to evaluate the fusion of classifiers. Particularly, the fusion of anomaly and signature-based ensemble learning will assist to improve the accuracy of deepfakes detectors.
- Existing research on deepfakes has mainly focused on detecting manipulation in the visual content of the video. However, audio manipulation, an integral component of deepfakes, is mostly ignored by the research community. There exists a need to develop unified deepfake detectors that are capable of effectively detecting both audio (i.e., TTS synthesis, voice conversion, cloned-replay) and visual forgeries (face-swap, lip-sync, and puppet-master) simultaneously.
- Existing deepfakes datasets lack the potential attributes (i.e. multiple visual and audio forgeries, etc.) required to evaluate the performance of more robust deepfake detection methods. The research community has hardly explored the fact that deepfake videos contain not only visual forgeries but audio manipulation as well. Existing deepfake datasets do not consider audio forgery and only focus on visual forgeries. In near future, the role of voice cloning (TTS synthesis, VC) and replay spoofing may increase in deepfake video generation. Additionally, shallow audio forgeries can easily be fused along-with deep audio forgeries in deepfake videos. We have already developed a voice spoofing detection corpus [311] for single- and multi-order replay attacks. Currently, we are working on developing a robust voice cloning and audio-visual deepfake dataset that can be effectively used to evaluate the performance of futuristic audio-visual deepfake detection methods.
- A unified method to address the variation of cloned attacks, such as cloned replay. The majority of voice spoofing detectors target detecting either replay or cloning attacks [159-161, 196]. These two-class oriented, genuine vs. spoof countermeasures, are not ready to counter multiple spoofing attacks on automatic speaker verification (ASV) systems. A study on presentation attack detection indicated that the countermeasures trained on a specific type of spoofing attack hardly generalizes well for other types of spoofing attacks [312]. Moreover, there does not exist a unified countermeasure that can detect replay and cloning attacks in multi-hop scenarios, where multiple microphones and smart speakers are chained together. We addressed the problem of spoofing attack detection on multi-hop scenarios in our prior work [10], but only for voice replay attacks. Therefore, there exists an urgent need to develop a unified countermeasure that can effectively detect a variety of spoofing attacks (i.e. replay, cloning, and cloned replay) in a multi-hop scenario.
- The exponential growth of smart speakers and other voice-enabled devices considers ASV a fundamental component. However, optimal utilization of ASV in critical domains, such as financial services, health care, etc., is not possible unless we counter the threats of multiple voice spoofing attacks on the ASV. Thus, this vulnerability also presents a need to develop a robust and unified spoofing countermeasure.
- There exists a crucial need to implement federated, learning-based, lightweight approaches to detect the manipulation at the source, so an attack doesn't traverse a network of smart speakers (or other IoT devices) [9,10].

# 7 Conclusion

This survey paper presents a comprehensive review of existing deepfake generation and detection methods. Not all digital manipulations are harmful. However, due to immense technological advancements, it is now very easy to produce realistic fabricated content. Therefore, malicious users can use it to spread disinformation to attack individuals and cause social, psychological, religious, mental, and political stress. In the future, we imagine seeing the results of fabricated content in many other modalities and industries. There is a cold war between deepfake generation and detection methods. As there are improvements in one it causes challenges for the other. We provided a detailed analysis of existing audio and video deepfake generation and detection techniques, along with their strengths and weaknesses. We have also discussed existing challenges and the future directions of both deepfake creation and identification methods.

# Acknowledgement

This material is based upon work supported by the National Science Foundation (NSF) under Grant numbers 1815724 and 1816019. Any opinions, findings, and conclusions or recommendations expressed in this material are those of the author(s) and do not necessarily reflect the views of the NSF.